\documentclass[journal]{IEEEtran}

\usepackage{cite}
\usepackage{amsmath,amssymb,amsfonts}
\usepackage{algorithmic}
\usepackage{graphicx}
\usepackage{textcomp}
\usepackage{xcolor}

\usepackage[ruled,vlined]{algorithm2e}
\usepackage{mathtools}  
\usepackage{pifont}
\usepackage{listings} 
\usepackage{tabularray}
\usepackage{color,soul}
\usepackage{array}
\usepackage{multirow}
\usepackage{ragged2e}
\usepackage{scalerel}
\usepackage{tikz}
\usepackage{balance}
\usepackage{xfrac}
\usetikzlibrary{svg.path}
\usepackage[hyphens]{url}
\usepackage[colorlinks=true,linkcolor=black,anchorcolor=black,citecolor=black,filecolor=black,menucolor=black,runcolor=black,urlcolor=black]{hyperref}

\definecolor{orcidlogocol}{HTML}{A6CE39}
\tikzset{
  orcidlogo/.pic={
    \fill[orcidlogocol] svg{M256,128c0,70.7-57.3,128-128,128C57.3,256,0,198.7,0,128C0,57.3,57.3,0,128,0C198.7,0,256,57.3,256,128z};
    \fill[white] svg{M86.3,186.2H70.9V79.1h15.4v48.4V186.2z}
                 svg{M108.9,79.1h41.6c39.6,0,57,28.3,57,53.6c0,27.5-21.5,53.6-56.8,53.6h-41.8V79.1z M124.3,172.4h24.5c34.9,0,42.9-26.5,42.9-39.7c0-21.5-13.7-39.7-43.7-39.7h-23.7V172.4z}
                 svg{M88.7,56.8c0,5.5-4.5,10.1-10.1,10.1c-5.6,0-10.1-4.6-10.1-10.1c0-5.6,4.5-10.1,10.1-10.1C84.2,46.7,88.7,51.3,88.7,56.8z};
  }
}

\newcommand\orcidicon[1]{\href{https://orcid.org/#1}{\mbox{\scalerel*{
\begin{tikzpicture}[yscale=-1,transform shape]
\pic{orcidlogo};
\end{tikzpicture}
}{|}}}}

\def\BibTeX{{\rm B\kern-.05em{\sc i\kern-.025em b}\kern-.08em
    T\kern-.1667em\lower.7ex\hbox{E}\kern-.125emX}}
\begin{document} 

\title{SHIFT: Dynamic Compute Relocation Framework for Communication-Aware Chiplet-Based Systems}

\author{
Arvin Delavari~\textsuperscript{\orcidicon{0009-0006-6350-0055}},~\IEEEmembership{Student Member,~IEEE,} 
Leonid Popryho~\textsuperscript{\orcidicon{0009-0002-0578-9592}},~\IEEEmembership{Graduate Student Member,~IEEE,}

Sneha Swaroopa~\textsuperscript{\orcidicon{0009-0000-7416-5316}},~\IEEEmembership{Graduate Student Member,~IEEE,}
Nader Sehatbakhsh~\textsuperscript{\orcidicon{0000-0001-7181-2258}},~\IEEEmembership{Member,~IEEE,}

Inna Partin-Vaisband~\textsuperscript{\orcidicon{0000-0002-6399-6672}},~\IEEEmembership{Senior Member,~IEEE,} 
and Boris Vaisband~\textsuperscript{\orcidicon{0000-0002-6176-5918}},~\IEEEmembership{Senior Member,~IEEE}
\vspace{-8mm}

\thanks{Arvin Delavari and Boris Vaisband are with the Department of Electrical Engineering and Computer Science, University of California, Irvine, Irvine, CA 92697 USA (e-mails: delavari@uci.edu, boris.vaisband@uci.edu).} 
\thanks{Leonid Popryho and Inna Partin-Vaisband are with the Department of Electrical and Computer Engineering, University of Illinois Chicago, Chicago, IL 60607 USA (emails: lpopry2@uic.edu, vaisband@uic.edu).} 
\thanks{Sneha Swaroopa and Nader Sehatbakhsh are with the Department of Electrical and Computer Engineering, University of California Los Angeles, Los Angeles, CA 90095 USA (emails: swaroopasneha25@g.ucla.edu,\linebreak nsehat@ucla.edu).} 
\thanks{This research is supported in part by the National Science Foundation (NSF) under Grant No. 2543560.} 
%\thanks{Corresponding author: Arvin Delavari}
}

%\markboth{The IEEE TRANSACTIONS ON COMPUTER-AIDED DESIGN OF INTEGRATED CIRCUITS AND SYSTEMS}%
%{A. Delavari \textit{et. al.}: SHIFT: Dynamic Compute Relocation Framework for Communication-Aware Chiplet-Based Systems}

%\IEEEpubid{0000--0000/00\$00.00~\copyright~20XX IEEE}
%\IEEEspecialpapernotice{(Invited Paper)}

\maketitle

\begin{abstract}

The increasing communication complexity of large-scale heterogeneous systems has motivated runtime methodologies for communication-aware workload placement and routing optimization. 
These communication limitations are addressed in this paper by proposing SHIFT—a novel runtime, topology-agnostic approach that transfers compute node context and data to a more suitably positioned node, rather than only shifting data, as in conventional networks-on-chip. 
The proposed strategy is evaluated on a wafer-scale chiplet-based architecture, utilizing a fine-pitch integration platform, featuring multiple bandwidth-domains for heterogeneous workloads. The proposed architecture employs multi-layered routing between functional or memory chiplets, and utility chiplets, which serve as intelligent nodes responsible for routing and the compute relocation framework. 
The adaptive scheduling and routing utilized a modified shortest-path algorithm for large-scale systems, complemented by a lightweight ML-assisted policy that infers traffic conditions to improve adaptivity. 
To establish a performance baseline, the initial assessment uses random instruction vectors and data patterns to evaluate the fundamental capabilities of SHIFT. Simulation results exhibit successful relocations over total trials ranging from 75.2\% to 97.9\% across configurations, with average latency improvements of 16.4\%–62.5\% and a maximum of 76.8\%. 
In addition, throughput is improved by up to 12.5$\times$,  power dissipation per unit area is reduced from 0.27~W/mm\textsuperscript{2} to 0.25~W/mm\textsuperscript{2} ($\sim$8\%), energy-per-bit is reduced by up to 58.3\%, and performance is improved by 18\% up to 149.7~PFLOPS. 
To evaluate efficiency under high logic and data density, the framework was tested on standard LLM workloads. 
Results exhibit average improvements of 4.9$\times$, 5.9$\times$, and 1.8$\times$ in, respectively, runtime, throughput, and energy-efficiency, while surpassing state-of-the-art wafer-scale LLM services and demonstrating strong compatibility with large-scale platforms and applications. 

\end{abstract}


\maketitle

\section{Introduction}
\label{introduction}

Communication overhead is a primary determinant of performance in large-scale computing systems. As demand grows for data-intensive applications--dominated by memory-compute operations, such as neural network (NN) training, machine learning (ML) inference, large language models (LLMs), automotive and robotics, and scientific computing--the cost of data movement is emerging as a critical bottleneck \cite{tsmc-1, Iyer_2019_ibm_journal}. 
Moreover, modern data-driven applications exhibit distinct workload asymmetry. For instance, in LLMs, the prefill stage is compute- and memory-bound, whereas the decode stage is memory-bandwidth-bound, leading to persistent resource underutilization across execution phases \cite{attention_paper, Azure_Splitwise}.

From a design automation perspective, emerging wafer-scale and heterogeneous architectures present complex optimization challenges, requiring runtime methodologies that jointly orchestrate workload placement, resource allocation, adaptive routing, and memory locality under dynamic workloads. While advanced packaging \cite{Lin_2020_jssc, Gomes_2020_isscc} and integrated high-bandwidth memory (HBM) stacks \cite{O’Connor_2017_micro} mitigate costs, efficient data orchestration and runtime scheduling remain significant hurdles. Excessive routing complexity, decision-making overhead, and long-range communication increase latency, degrade signal integrity, and stifle system-level efficiency, necessitating a holistic optimization framework that treats runtime data movement as a first-class design constraint for bridging physical-level integration and system-level communication. 

A novel compute relocation framework, \textbf{SHIFT}, addressing communication and performance limitations in large-scale systems, is proposed in this paper. Unlike conventional approaches where data is moved from shared memory to compute chiplets, \textbf{SHIFT is a runtime communication-aware routing framework that dynamically relocates execution to communication-optimal compute nodes, reducing communication costs without imposing additional computational overhead on functional cores for adaptive routing and scheduling.}
The proposed approach is topology-agnostic, yet this work focuses on chiplet-based systems, which are better suited for large-scale platforms and LLMs as the target application. 
The main contributions of this paper are: 

\begin{itemize}

\item A runtime framework is proposed for communication-aware relocation of compute nodes and data to strategically-positioned nodes with lower communication costs compared to networks-on-chip (NoCs). 

\item A custom optimized shortest-path function and an ML-assisted variant are developed and evaluated on multiple router core architectures for adaptive inter-chiplet routing, yielding up to a 22$\times$ runtime improvement. 

\item A heterogeneous network architecture is introduced with virtually-stacked multi-layer routing, comprising high-bandwidth domains (HBDs) and general-purpose domains (GPDs) for asymmetric workload requirements.

\item The proposed framework is evaluated over a fine-pitch integration platform, demonstrating up to 12.5$\times$ improvement in throughput and 58.3\% in energy-efficiency. 

\item SHIFT is further evaluated on standard LLM workloads achieving average 4.9$\times$, 5.9$\times$, and 1.8$\times$ improvements in runtime, throughput, and energy-efficiency, outperforming state-of-the-art (SOTA) wafer-scale and chiplet-based LLM services and compute platforms. 
    
\end{itemize} 

The rest of this paper is organized as follows. Background and motivation are provided in Section~\ref{background}, while the network-on-interconnect fabric (NoIF) platform utilized for the experiments is introduced in Section~\ref{noif-section}. The proposed framework is detailed in Section~\ref{proposed-framework}, and the experimental setup, LLM benchmark evaluations, and SOTA comparisons are presented in Section~\ref{evaluations}, followed by concluding remarks in Section~\ref{conclusions}.

\section{Background and Motivation}\label{background} 

Large-scale applications demand cross-layer co-design across software, architecture, and physical layers. This section reviews SOTA communication-aware scheduling and design space exploration (DSE) techniques, alongside the benefits of advanced packaging for AI hardware, to contextualize the proposed framework.

\textbf{Scheduling and DSE approaches:} Traditional process migration for load balancing incurs excessive latency from full-state transfers, making it suboptimal for dynamic AI workloads~\cite{Tan_ieee_tcc_2025}. In NoCs, task reassignment is a reactive mechanism to resolve deadlocks or alleviate congestion by remapping tasks to underutilized nodes~\cite{Lei_dac_2017}. While foundational, these methods lack the global optimization required for dynamic workload-aware chiplet systems. Furthermore, while static scheduling (\textit{e.g.}, Groq~\cite{groq_paper}) or predictive heuristics~\cite{Feldmann_micro_2024} reduce routing complexity, they remain insufficiently adaptable to volatile traffic patterns, highlighting the need for a communication-aware, runtime-adaptive scheduling and resource allocation frameworks.

\textbf{Processing-in-Memory (PIM) and Processing-near-Memory (PNM):} Another approach used for enhancing memory-compute proximity and reducing routing overhead is PIM. This paradigm integrates simple computational logic directly within or near memory arrays, allowing basic and bitwise operations to be executed without transferring data to compute cores \cite{Seshadri_micro_2017, GDDR6_PIM}. This approach reduces latency and energy consumption, especially for graph processing or sparse matrices \cite{Mamdouh_2025_tcad, PIMSAB_paper}. 
Embedding logic within dense memory arrays is challenging, with thermal and area limits providing only limited support for general-purpose workloads. This in-situ operation further prevents efficient scaling in systems demanding programmability and heterogeneous processing \cite{SpecPIM}, making it unsuitable for large-scale architectures. 

PNM has emerged as a more realistic and widely adopted solution in real‑world implementations. Bringing memory closer to--or ideally within--the package enables lower‑latency communication. Advanced packaging further enable high‑bandwidth, low‑parasitic, high‑speed communication. Fine‑pitch near‑memory links complement the proposed approach through locality‑driven compute relocation \cite{Gomes_2020_isscc}. 

\textbf{Chiplet-based wafer-scale integration:}  
Recent advances in packaging are shifting chiplet integration from traditional interposer-based substrates to fine‑pitch integration platforms~\cite{Iyer_2019_ibm_journal, Lin_2020_jssc, Hu_2023_ectc}. By enabling tighter chiplet spacing and higher shoreline utilization, fine‑pitch integration reduces link parasitics while increasing bandwidth. In contrast, interposer-based packaging supports fewer chiplets and relies on serializer/deserializer (SerDes) links, limiting scalable, adaptable, and efficient communication. 

Fine‑pitch integration reduces inter‑chiplet distances, enabling high‑bandwidth parallel communication and accommodating more chiplets per substrate. Consequently, \textbf{wafer‑scale chiplet-based systems} leveraging heterogeneous integration can serve as an alternative to monolithic wafer‑scale platforms.

\section{Network on Interconnect Fabric (NoIF)}
\label{noif-section}

\textbf{Motivation:} Novel interposer-aware topologies and layouts improve latency, throughput, and bisection bandwidth by exploiting domain distinction and optimized placement~\cite{Wang_2025_tcad, amd_dac_2020, Chen_2024_ieee_tcad}. These methods, however, are confined to interposer-based systems and fail to scale to wafer-level platforms, where long- and mid-range communication remains limited by packaging, topology, and floorplanning. Prior studies on memory-compute co-design~\cite{Kou_2025_iccad} highlight that efficient communication is still critical to meet modern bandwidth demands~\cite{WaferLLM_2025, Xu_isca_2025_llm}. 

To overcome limitations in wafer-scale architectures and packaging constraints, we utilize the network-on-interconnect-fabric (NoIF) concept--a logical 3D-stacked hybrid architecture \cite{Safari_2024_ectc} optimized for hierarchical communication--as a key system-level enhancement in ultra-large-scale systems. 

\begin{figure}
    \centering
    \includegraphics[width=\linewidth]{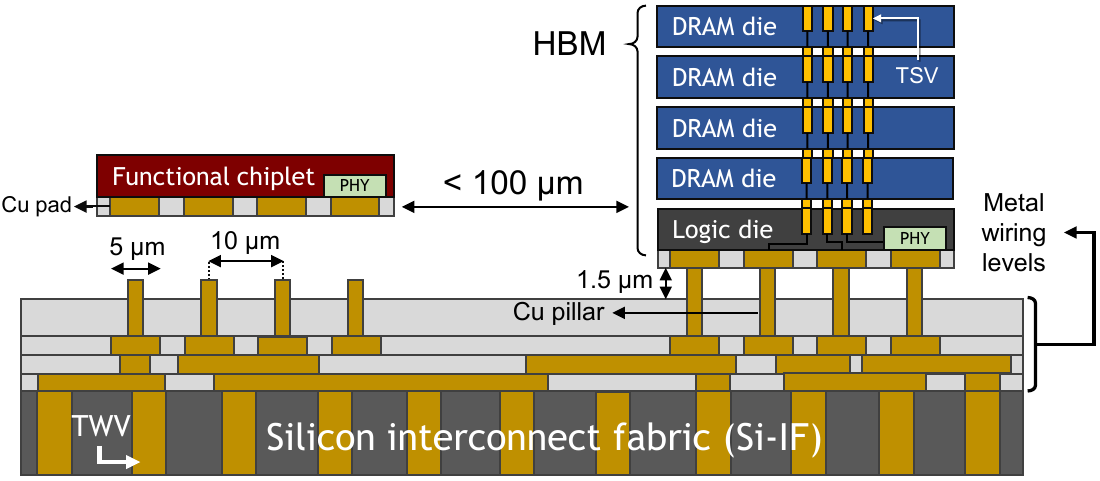}
    \vspace{-6.5mm}
    \caption{Chiplet-based integration of a functional chiplet (FC) and a fine-pitch high-bandwidth memory (HBM) chiplet on silicon interconnect fabric (Si-IF).}
    \label{fig:si-if}
\end{figure}

\textbf{Packaging:} 
The  packaging technology in this work, the silicon interconnect fabric (Si-IF), provides fine-pitch ($\le$10~\textmu m) and metal–metal thermal compression bonding (TCB) vertical links and short ($\le$100~\textmu m) horizontal links on silicon, enabling multi-terabit communication with low latency and power, while eliminating the need for interposers or interconnect bridges. Chiplet-based integration on Si-IF is shown in Fig.~\ref{fig:si-if}.

\begin{figure*}
    \centering
    \includegraphics[width=0.9\linewidth]{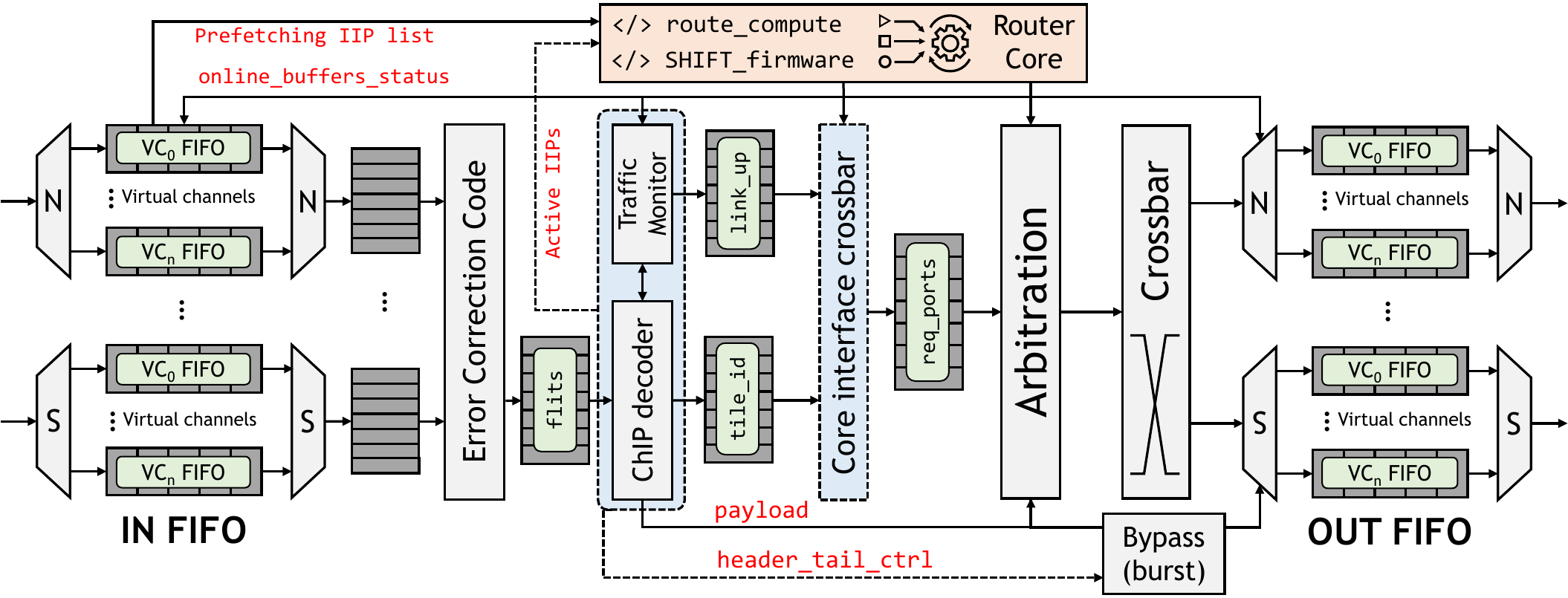}
    \vspace{-2.4mm}
    \caption{Microarchitecture of the UC multi-range router \cite{Delavari_2026_glsvlsi}, featuring a dedicated processor core that replaces the conventional, deterministic route compute module in the NoC router pipeline to support dynamic communication- and workload-aware compute relocation, congestion-awareness, and adaptive routing.}
    \label{fig:utility-chiplet-router}
    \vspace{-4mm}
\end{figure*}

\textbf{Architecture:} The proposed system integrates utility chiplets (UCs) as intelligent nodes managing communication, power delivery, synchronization, inter-chiplet control, and testing across the substrate~\cite{Safari_iscas_2021, Vaisband_2019_slip, Delavari_2026_glsvlsi}. UCs handle long-haul packetized communication, congestion awareness, rerouting, and the communication-aware relocation strategy. The proposed UC router includes a processing core for adaptive and runtime routing and relocation decision. Prefetching packets metadata enables this core to shorten decision paths. The UC router microarchitecture is illustrated in Fig.~\ref{fig:utility-chiplet-router}. 

\begin{figure}
    \centering
    \includegraphics[width=\linewidth]{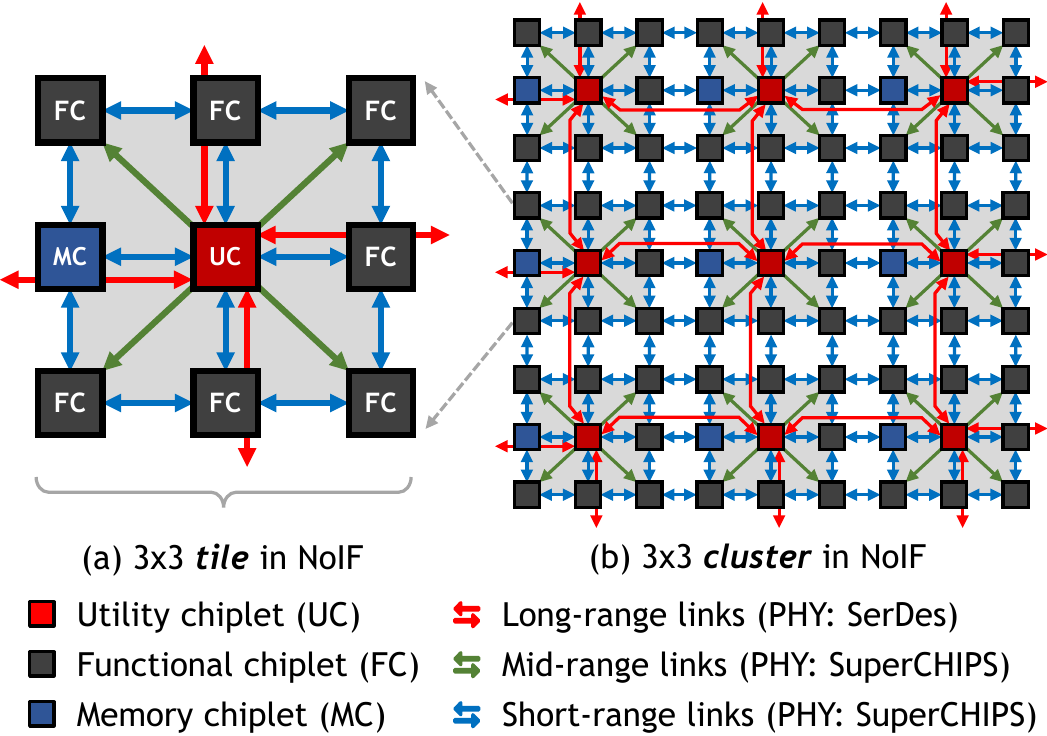}
    \vspace{-6mm}
    \caption{A hierarchical view of a 3\texttimes3 tile of a 3\texttimes3 cluster arrangement with multi-range routing in the proposed NoIF topology.}
    \label{fig:tile-cluster}
\end{figure}

Functional chiplets (FCs) serve as firmware execution nodes (\textit{i.e.}, CPUs, GPUs, or inference engines), while memory chiplets (MCs) provide distributed memory and may use HBM, SRAM, or DRAM; in our experiments, MCs are HBM4 chiplets. In the proposed system, a \textbf{tile} consists of nine chiplets in a 3\texttimes3 mesh: seven FCs, one HBM (MC), and a central UC. A \textbf{cluster} forms a 2D grid of tiles enabling scalable wafer‑scale integration. All chiplets support short‑range parallel communication, while UCs also handle mid‑ and long‑range communication (see \textbf{Topology}). An example tile–cluster configuration is shown in Fig.~\ref{fig:tile-cluster}.

\begin{figure}
    \centering
    \includegraphics[width=\linewidth]{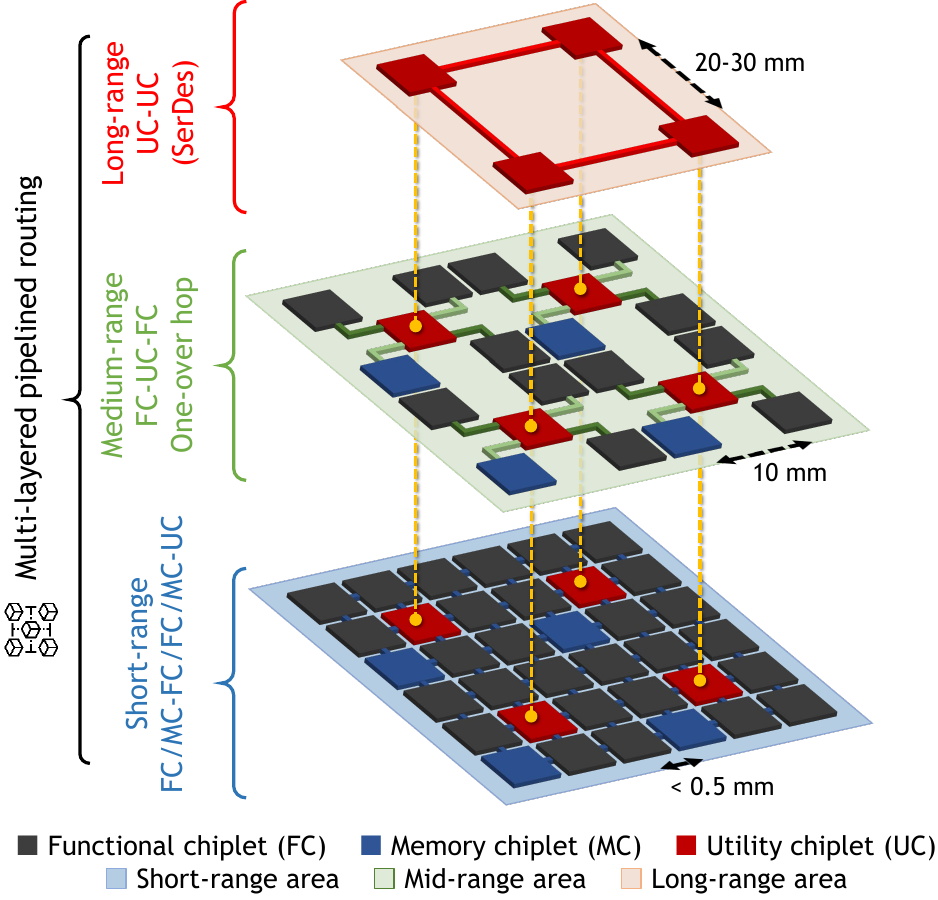}
    \vspace{-6.75mm}
    \caption{Physical communication ranges and routing layers in the proposed hybrid communication infrastructure with a virtually-3D-stacked topology~\cite{Delavari_2026_glsvlsi}.}
    \label{fig:3d-topology}
\end{figure}

\textbf{Topology:} The NoIF tiles and clusters utilize a hybrid multi-range topology based on the chiplet interface protocol (ChIP), designed for advanced packaging. This approach supports both serial and parallel chiplet communication, overcoming the bandwidth limitations of SOTA alternatives like UCIe~\cite{Delavari_2025_ieee_jetcas}. In this bi-directional communication, mid-range one-hop diagonal transfers occur through UCs via repeaters and bypass channel, and long-range UC-to-UC SerDes links enable system-wide high-bandwidth communication, as illustrated in the virtually 3D-stacked arrangement in Fig.~\ref{fig:3d-topology}. Hierarchical routing can assist the router computational core to reduce congestion by using layer-specific algorithms. 

\textbf{Summary:} By leveraging the channel characteristics of fine-pitch integration platforms together with network/architectural features, the NoIF enables scalable heterogeneous chiplet integration, suitable for ultra-large-scale applications. 
% A parametric speedup comparison against conventional 2D-mesh topology is illustrated in Fig.~\ref{fig:noif-speedup}.

% \begin{figure}
%     \centering
%     \includegraphics[width=\linewidth]{figures/noif_speedup.pdf}
%     \vspace{-7mm}
%     \caption{NoIF vs. 2D-mesh normalized latency and speedup against manhattan hop distance in the network.}
%     \label{fig:noif-speedup}
% \end{figure}

\section{SHIFT Framework}
\label{proposed-framework}

\begin{figure}
    \centering
    \includegraphics[width=\linewidth]{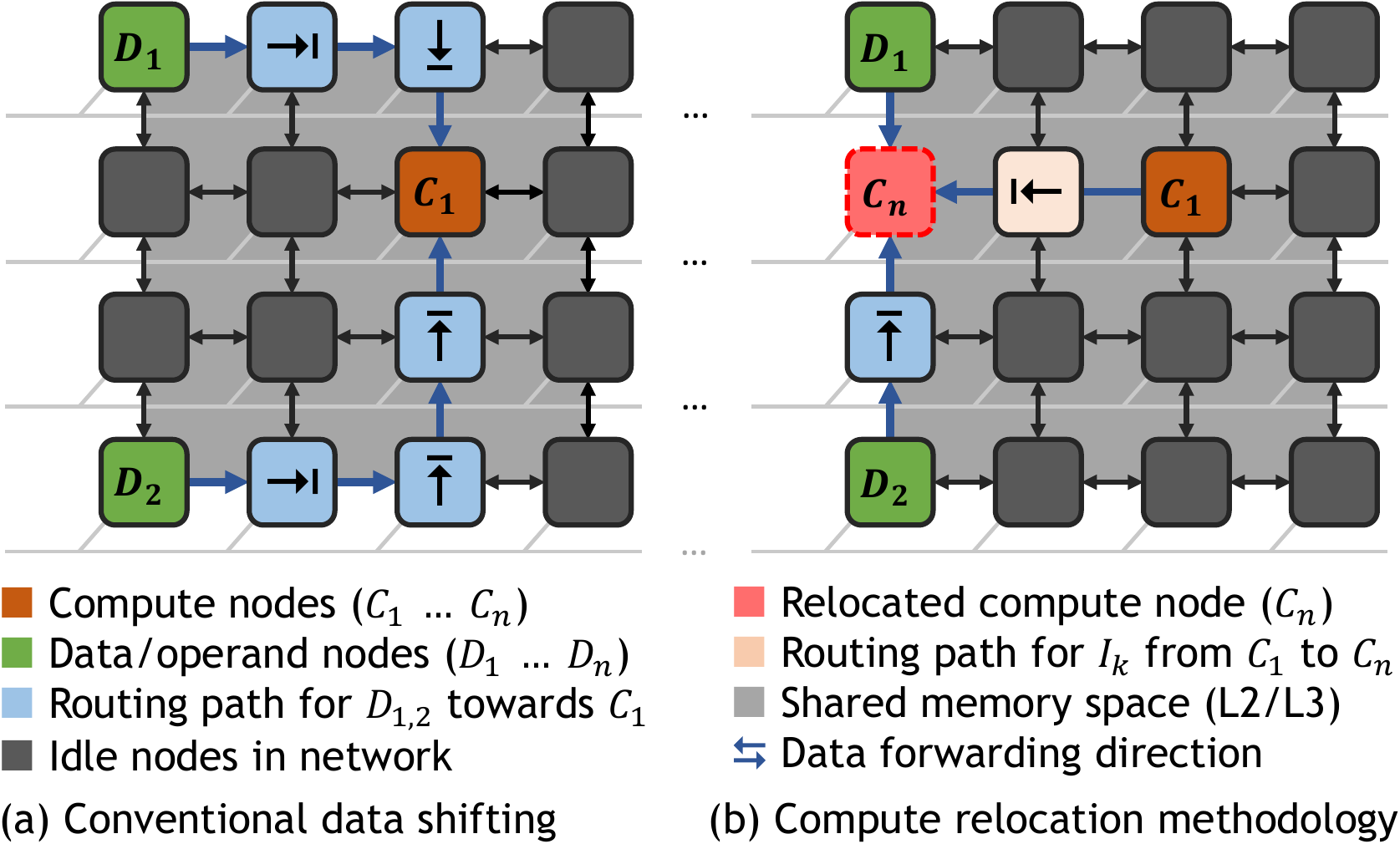}
    \vspace{-6.75mm} 
    \caption{Comparison between the dataflow of conventional data shifting mechanism in NoCs and compute relocation in a shared-memory system with mesh arrangement, for a single instruction.} 
    \label{fig:example}
\end{figure}

In this section, the proposed system-level optimization framework is presented, along with considerations for enabling this methodology on wafer-scale platforms. An example highlighting the differences between conventional NoC routing and communication-aware dynamic relocation, is shown in Fig.~\ref{fig:example}. The example is based on a basic mesh configuration, as the proposed strategy is \textbf{topology-agnostic and independent of the underlying platform}. 

In Fig.~\ref{fig:example}~(a) (conventional XY routing), operands $S=\{D_1, D_2\}$ are routed through intermediate nodes $V$ to the initially-scheduled compute chiplet $C_1 \in V$. 
In Fig.~\ref{fig:example}~(b) (compute relocation), a node $C_n \in V$ is selected to minimize total communication cost plus controller overhead. We define the execution cost at $C_n$ as:
\begin{equation}\small
    J(C_n) = \sum_{D_i \in S} \text{dist}(D_i, C_n) + \text{dist}(C_1, C_n)+ \tau(C_1, C_n\in V)
\end{equation}
where $\text{dist}(\cdot)$ represents the communication cost and $\tau(\cdot)$ is the relocation decision and dispatching overhead. Relocation is performed if $\min_{C_n \in V} J(C_n) < \sum_{D_i \in S} \text{dist}(D_i, C_1)$, effectively minimizing routing overhead by relocation to $C_n$.

\subsection{Considerations and Design Methodology}
\label{design-space}

The SHIFT framework is orchestrated by the core in UCs. As outlined in Section~\ref{noif-section}, UCs function as network intelligence nodes rather than executing firmware-related instructions. UCs monitor and analyze communication patterns, handle mid- and long-range routing, and coordinate relocation. 

\subsubsection{Framework Workload Balancing}
\label{framework-workload-balancing}

A key design consideration is minimizing the \textbf{performance impact of the relocation framework} itself. In the proposed strategy, routing and relocation decisions are offloaded to UCs, ensuring that FCs remain dedicated to firmware-related tasks with minimal interruption. This decoupling strategy enables efficient workload distribution and runtime resource reallocation, as framework overhead does not consume valuable compute resources. 

The relocation decision compares the total communication latency of keeping a task at its current node $i$ versus relocating it to a candidate node $j$ using a predefined network map and communication cost models. A task contains an instruction and data chunks ${D_k}$ distributed across nodes. The communication latency between nodes $a$ and $b$ for a message depends on the network latency and the link bandwidth. 
In the baseline case, execution remains at node $i$, requiring all data chunks ${D_k}$ from nodes $k$ to be transferred to $i$, so the total latency equals the maximum of communication latencies from each node $k$ to $i$. 
In the proposed framework, the task metadata is first sent to the UC, which decides whether execution remains at $i$ or relocates to node $j$. If relocation is selected, the instruction is transferred from $i$ to $j$, and all data chunks ${D_k}$ are transferred from nodes $k$ to $j$. The resulting latency includes metadata handling overhead, instruction transfer latency, and the maximum of data-transfer latencies to node $j$. 
\textbf{Thus, SHIFT is beneficial only when the communication savings from improved data locality exceed the relocation and metadata overhead; otherwise, it can introduce additional latency and throughput degradation.} 

\subsubsection{Instruction Intent Packet (IIP)} 
\begin{figure}[t]
    \centering
    \includegraphics[width=\linewidth]{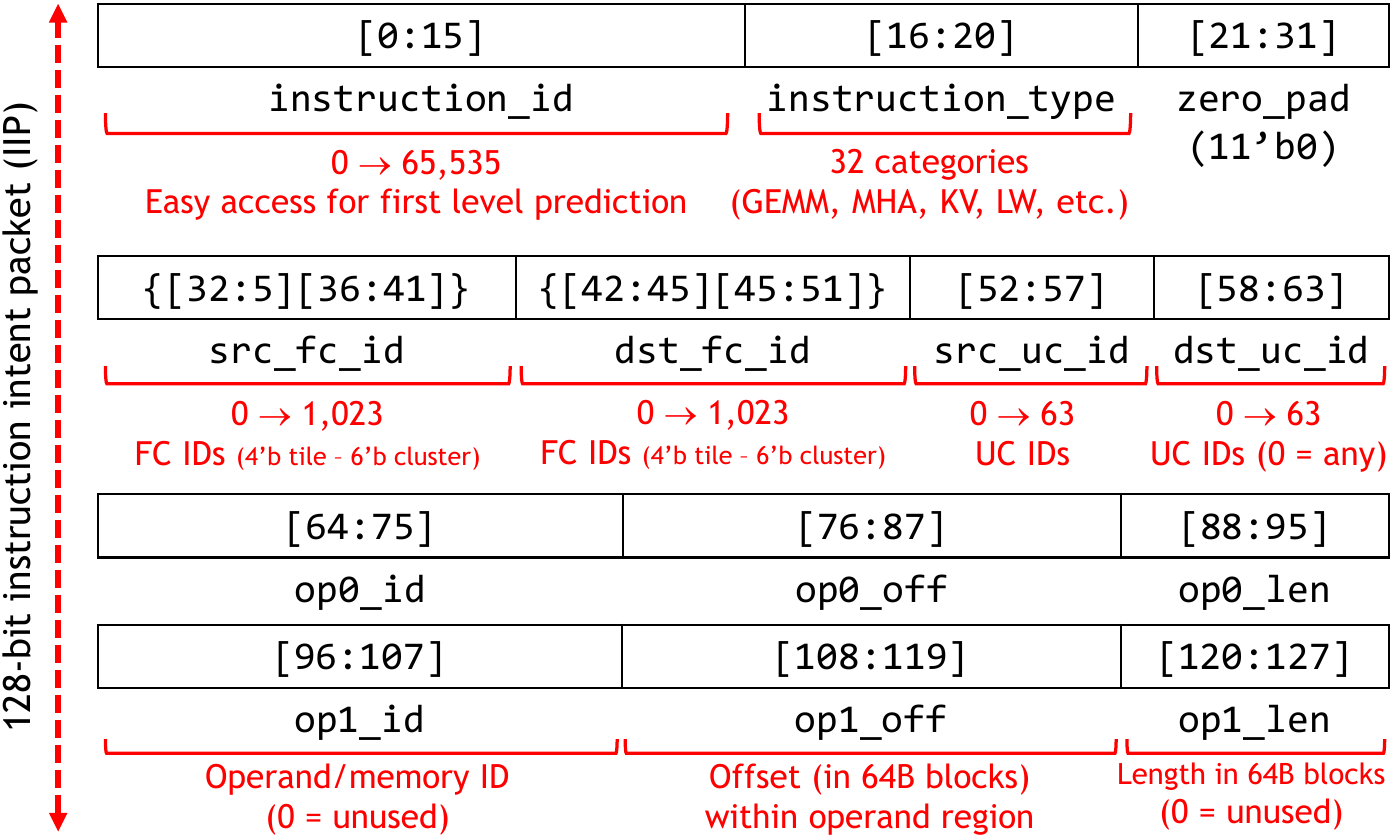}
    \vspace{-6.5mm}    
    \caption{Instruction intent packet (IIP) for instruction metadata transfer.}
    \label{fig:iip}
\end{figure}

As mentioned in Section~\ref{framework-workload-balancing}, the proposed method relies on \textbf{early instruction metadata propagation from FCs to UCs} at dispatch time, \textit{i.e.}, when an FC is ready to execute an instruction but before operand commitment. The FC sends a lightweight instruction intent packet (IIP) to its nearest UC, incurring minimal communication overhead, and the UCs execute the framework to determine relocation and routing. The IIP format is shown in Fig.~\ref{fig:iip}. The IIP is designed in a minimal format to fit all payload variations over short‑ and mid‑range links and to enable easy diagnosis and decoding by the ChIP \cite{Delavari_2025_ieee_jetcas, Delavari_2026_glsvlsi}.

\subsubsection{Congestion Awareness}
Each UC monitors buffer occupancy and link utilization, enabling low-overhead local traffic awareness in each tile. Inter-cluster UCs share summarized statistics, forming a \textbf{distributed traffic sensing} mesh without relying on a central controller. 
They periodically broadcast compressed traffic metrics and employ an \textbf{event-driven gossip protocol} (\textit{i.e.}, broadcasting only upon reaching congestion thresholds) for global traffic visibility. 

\subsection{Proposed Framework}
\label{proposed-framework-steps}

\begin{figure*}
    \centering
    \includegraphics[width=0.95\linewidth]{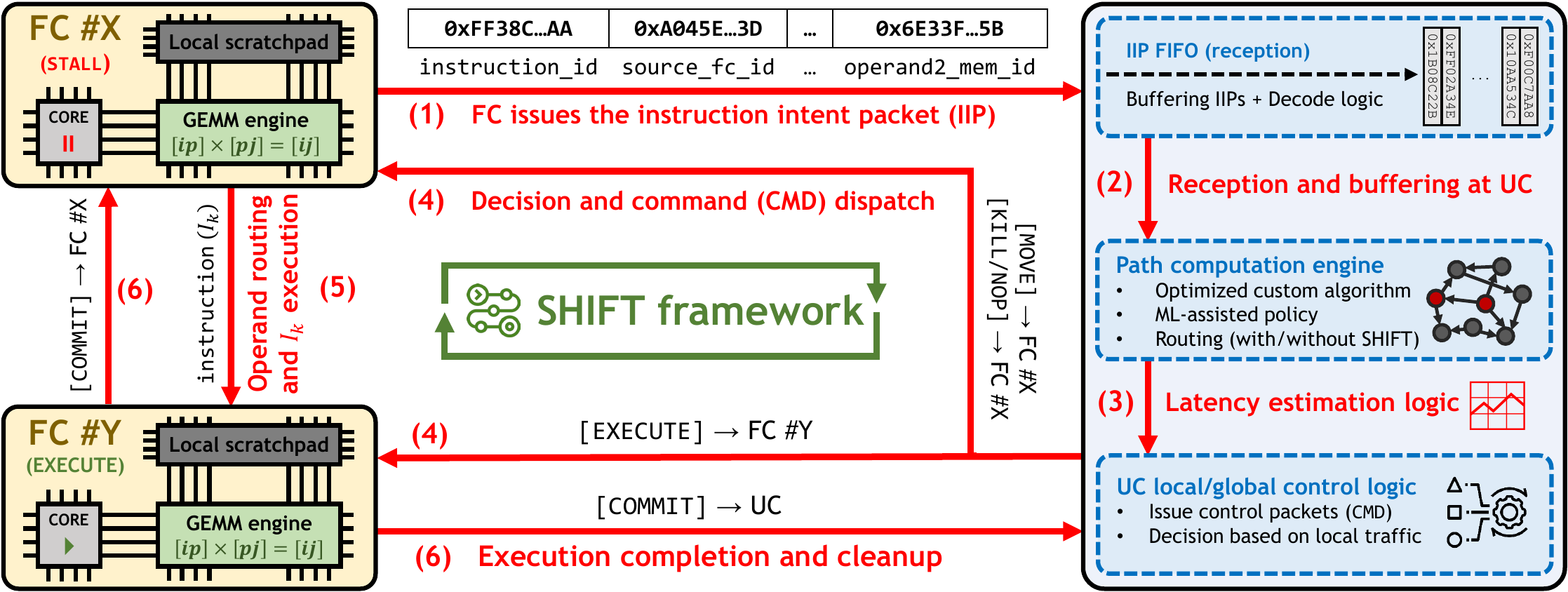}
    \vspace{-1.75mm}
    \caption{SHIFT compute relocation framework main stages: (1)~FC issues the instruction intent packet (IIP). (2)~Reception, buffering, and decode logic at UC. (3)~Execution of latency estimation algorithms at UC (shortest-path, \textit{e.g.}, Dijkstra, A*). (4)~Decision and command (CMD) dispatch. (5)~Operand routing and instruction execution (6)~Execution completion and cleanup.}
    \label{fig-step-by-step}
    \vspace{-2.75mm}
\end{figure*}

Given the established design space and policies in Section~\ref{design-space}, the SHIFT framework proceeds in six primary steps, which is illustrated in Fig.~\ref{fig-step-by-step} and explained in this section. 

\subsubsection{FC Issues the Packet} 
The executing FC generates and transmits a compact IIP to its nearest UC while retaining the full instruction and operands locally. After transmission, the FC enters a \texttt{STALL} state awaiting execution clearance. The packet includes: instruction ID, source/destination cluster/tile/chiplet IDs, instruction types, operand IDs, and address information (details in Fig.~\ref{fig:iip}). The routing is initiated by targeting the geographically closest (statically assigned) UC. 

\subsubsection{Reception and Buffering} 
Upon arrival, the IIP is buffered in a FIFO. Prefetching is enabled through UC core for idle-state decision-making. The path computation engine then locates the operands/kernels, identifies the source FC, and selects potential destination FCs based on local congestion, using the decode logic integrated within the reception stage. 

\subsubsection{Latency Estimation (Shortest-Path Algorithm)}
The rising demand for large-scale chiplet-based systems has driven the development of link-aware routing algorithms~\cite{Taheri_2024_tcad, Liu_2026_tcad}. This subsection presents our approach to meeting these system-level requirements for the proposed NoIF. 

\textbf{Shortest-path algorithm:} 
To determine the suitable relocation destination, each UC evaluates candidate FCs based on a predefine network map and communication cost models which also incurs additional computation overhead. 
The proposed routing policy is a custom Dijkstra-like~\cite{dijkstra_1, dijkstra_2} algorithm optimized for low-latency, scaled-out execution, shown in Algorithm~\ref{alg:bidirectional-dijkstra}. 
A local subgraph is formed from the decoded IIP, with \texttt{src\_node} and \texttt{dst\_node} at opposite edges to limit the bidirectional search space, which runs parallel from both nodes. The first intersection defines the \texttt{inter\_node}, minimizing latency while avoiding full-network traversal. 

\begin{algorithm}[t]
    \small
    \caption{Modified Shortest Path Algorithm}
    \label{alg:bidirectional-dijkstra}
    \KwData{Decoded IIP, \texttt{src\_node}, \texttt{dst\_node}, network graph $G$}
    
    \tcp{1: Subgraph formation}
    Derive subgraph $G_\text{sub}$ from $G$ with \texttt{src\_node} and \texttt{dst\_node} placed at opposite edges\;
    Initialize all nodes in $G_\text{sub}$ \;
    
    \tcp{2: Bidirectional initialization}
    Set $\text{dist}[\texttt{src\_node}]=0$ and $\text{dist}[\texttt{dst\_node}]=0$\;
    Initialize priority queues $Q_\texttt{src}$ and $Q_\texttt{dst}$ with \texttt{src\_node} and \texttt{dst\_node}\;

    \tcp{3: Parallel expansion}
    \While{$Q_\texttt{src}$ and $Q_\texttt{dst}$ not empty}{
        Expand one hop from \texttt{src\_node} side in $Q_\texttt{src}$\;
        Expand one hop from \texttt{dst\_node} side in $Q_\texttt{dst}$\;
        \If{a common node (\texttt{inter\_node}) is reached}{
            Record $\texttt{inter\_node}$ as intersection point\;
            \textbf{break}\;
        }
    }

    \tcp{4: New path construction}
    Combine partial paths from \texttt{src\_node} to \texttt{inter\_node} and \texttt{dst\_node} to \texttt{inter\_node}\;
    Compute final relocation path $\text{P}_\texttt{inter}$ through \texttt{inter\_node}\;
    Output $\text{P}_\texttt{inter}$ and estimated minimal latency $\text{L}_\texttt{inter}$\;
\end{algorithm}

\begin{figure}
    \centering
    \includegraphics[width=\linewidth]{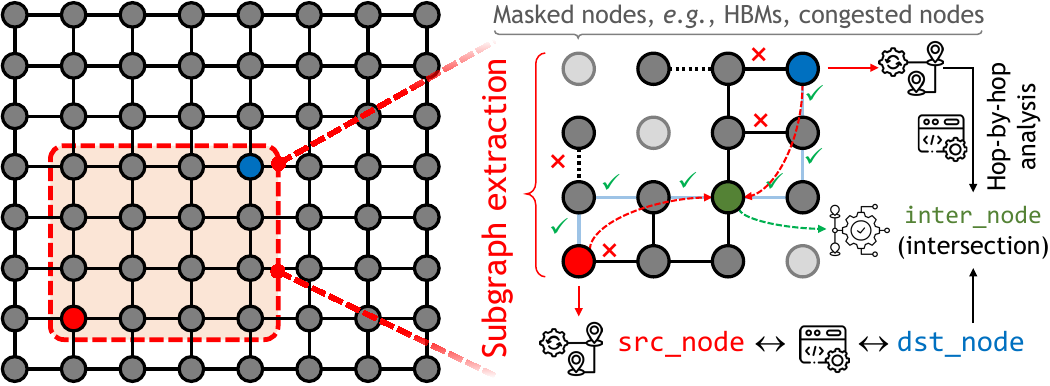}
    \vspace{-6.75mm} 
    \caption{High-level representation of the shortest-path function (Algorithm~\ref{alg:bidirectional-dijkstra}).} 
    \label{fig:algorithm}
\end{figure}

Since the topology deviates from a standard mesh, deterministic routing alone cannot meet application-level requirements. Subgraph extraction is employed solely to reduce exploration space, with MCs masked from the routing table to save computation. Although \textbf{not guaranteeing the optimal node}, this method identifies \textbf{better-positioned nodes} with lower end-to-end (E2E) latency for both routing and decision-making in large-scale networks. 
A high-level representation of the shortest-path function is provided in Fig.~\ref{fig:algorithm}.  
For intra-tile communication, the framework is mostly omitted, as it is more costly than standard dimension-ordered routing (DOR). 

Deadlocks may occur when nodes are removed from routing tables due to high buffer or link utilization thresholds. If this happens before the latest network status update, packets cannot be rerouted and are dropped. Although such cases are rare in application-driven traffic, simple deadlock-avoidance policies for routing retries can be applied to mitigate such issues. 

\textbf{ML-assisted routing:} 
In Algorithm~\ref{alg:bidirectional-dijkstra}, the time complexity is $O(|E_\text{sub}|\log|V_\text{sub}|)$, where $E$ and $V$ represent the numbers of edges and vertices, respectively. 
Since the algorithm relies on static edge weights, it is sensitive to transient congestion and may require recomputation. 
Prior studies show that some of the ML-based models~\cite{7229288, LPNet, Fast-and-Accurate-NoC-Latency-Estimation-for-Application-Specific-Traffics-via-Machine-Learning} predict latency more accurately and efficiently than heuristic approaches~\cite{PreNoc, DRLAR}. Consequently, the routing step can be replaced with an ML-driven, per-hop, congestion-aware predictor. 

For UC $u$, destination $dst$, and neighbor $v\!\in\!\mathcal{N}(u)$, a model $\hat{L}(u\!\to\!v\,|\,dst,\phi)$ is used to estimate the residual latency from $v$ to $dst$ based on local features $\phi$ (\textit{e.g.}, queue depths, link utilizations, gossip summaries). The next hop is selected as:
\begin{equation}
    v^\star=\arg\min_{v\in\mathcal{N}(u)} \hat{L}(u\!\to\!v\,|\,dst,\phi),
\end{equation}
\noindent after which the remaining SHIFT stages (IIP emission, dispatch, operand routing, commit) proceed unchanged. A per-hop decision cost of $O\!\big(\deg(u)\,C_{\mathrm{MLP}}\big)$ is incurred, where $C_{\mathrm{MLP}}$ denotes the forward-pass cost. Per packet, the cost is $O\!\big(\sum_{j=1}^{H}\deg(u_j)\,C_{\mathrm{MLP}}\big)$ for path length $H$ (or $O(\bar d\,H\,C_{\mathrm{MLP}})$ when degrees are approximately constant). 

\subsubsection{Decision and Command Dispatch}
If the source FC is optimal, the UC sends an $\texttt{EXECUTE}$ control packet to it. Otherwise, it sends a $\texttt{SHIFT\_TO}$ command to the source FC and corresponding MCs/FCs (data locations), and an \texttt{EXECUTE} command (with metadata) to the new destination FC. A $\texttt{KILL}$/$\texttt{NOP}$ command is then sent to the original FC to discard its local instruction copy. 

\subsubsection{Operand Routing and Instruction Execution} 
The selected FC prepares for execution by configuring input buffers and decoder units. UCs facilitate congestion-aware and burst-enabled routing \cite{Delavari_2025_ieee_jetcas}, within the network to transfer operands. 

\subsubsection{Execution Completion and Cleanup} 
After execution, results are written back to the target MCs. The destination FC sends a \texttt{COMMIT} packet to the UC, which removes the corresponding entry from its transaction table, indexed by \texttt{instruction\_id} (mapped to \texttt{TID} in ChIP controller). 

\subsection{Execution Flow}
\label{compute-relocation-D}

\subsubsection{Framework Execution} 

Algorithm~\ref{alg:packet-latency} describes the initial SHIFT stages, where the FC sends a compact IIP to the nearest UC, which buffers it and evaluates candidate FCs using the optimized shortest-path function (Algorithm~\ref{alg:bidirectional-dijkstra}). 
Algorithm~\ref{alg:dispatch-routing-exec} then handles relocation and execution: the UC issues control signals for local or shifted execution, starts once dependencies resolve, and upon completion, writes results to memory and issues a \texttt{COMMIT} to finalize and release resources. 

\begin{algorithm}[t]
    \small
    \caption{IIP Handling and Latency Estimation}
    \label{alg:packet-latency}
    \KwData{IIP from source\_FC, Candidate FCs}
    
    \While{TRUE}{
        \uIf{IIP issued by \texttt{source\_FC}}{
            Transmit IIP to nearest UC \\
            source\_FC $\leftarrow$ \texttt{STALL} state
        }
        \uIf{UC receives IIP}{
            Buffer IIP \\
            \ForEach{candidate FC\_X}{
                Compute costs:
                \[
                \begin{cases}
                \text{C}_\text{1} = \text{short\_path}(\text{operand1\_mem} \to \text{FC}\_\text{X}) \\
                \text{C}_\text{2} = \text{short\_path}(\text{operand2\_mem} \to \text{FC}\_\text{X}) \\
                \text{C}_\text{3} = 0, \textbf{ if } \text{FC}\_\text{X} = \text{source\_FC}; \textbf{ else }\\ 
                ~~~~~~~~~~~\text{short\_path}(\text{source\_FC} \to \text{FC}\_\text{X}) \\
                \text{C}_{\text{total}} = \max(\text{C}_\text{1}, \text{C}_\text{2}, \text{C}_\text{3}) + \text{comp\_overhead}
                \end{cases}
                \]
            }
            Select FC\_X with minimum $\text{C}_{\text{total}}$ \\
            \Return selected\_FC\_X
        }
    }
\end{algorithm}

\begin{algorithm}[t]
    \small
    \caption{Decision Dispatch and Execution}
    \label{alg:dispatch-routing-exec}

    \KwData{selected\_FC\_X, source\_FC, IIP, UC buffer}
    
    \If{selected\_FC\_X = source\_FC}{
        send (\texttt{EXECUTE}, source\_FC)
    }
    \Else{
        send (\texttt{SHIFT\_TO}, source\_FC) \\
        send (\texttt{EXECUTE}, selected\_FC\_X, metadata) \\
        send (\texttt{KILL}, source\_FC)
    }
    
    Prepare selected\_FC input buffers and decoder \\
    UC routes operands using congestion-aware routing \\
    
    \If{Operands received and FC ready}{
        Execute instruction
    }
    
    \If{Execution done}{
        Writeback results to memory \\
        send (\texttt{COMMIT}, UC) \\
        UC removes \texttt{TID} from transaction table \\
        FC exits \texttt{STALL} state
    }
\end{algorithm}

\subsubsection{Memory-Aware Predictive Policies} 

In scenarios where communication costs do not outweigh the latency of relocation, the proposed framework may \textbf{introduce additional stalls} compared to conventional routing. Just as branch prediction can improve execution decisions in a core, poor relocation choices can degrade performance. To mitigate mispredictions, the platform incorporates a memory-aware predictive policy, illustrated in Algorithm~\ref{alg:prediction-policy}. 

\begin{algorithm}[t]
    \small
    \caption{Memory-Aware Relocation Prediction}
    \label{alg:prediction-policy}
    \KwData{
    IIP, source\_FC, Memory domain map, Threshold (relocation\_gain\_margin)
    }

    \uIf{(operand1\_mem AND/OR operand2\_mem) $\in$ L1 of source\_FC}{
        \Return Execute locally (Relocation skipped) \\
        Mark instruction as: \texttt{LOCAL\_PREFERRED}
    }
    \uElseIf{operand1\_mem AND/OR operand2\_mem $\in$ Tile's MC}{
        \Return Execute Locally (Prefetch nearby) \\
        Mark instruction as: \texttt{LOCAL\_PREFERRED}
    }
    \Else{
        Evaluate Relocation\\
        Compare $\text{C}_{\text{SHIFT}}$ vs. $\text{C}_{\text{base}}$
        
        \uIf{$\text{C}_{\text{SHIFT}} < \text{C}_{\text{base}} - \text{relocation\_gain\_margin}$}{
            \Return Proceed with SHIFT
        }
        \Else{
            \Return Execute locally (SHIFT rejected)
        }
    }
\end{algorithm}

This mechanism evaluates whether the operand addresses are already present in the tile-based shared memory of the source FC location. If so, relocation is bypassed, and the instruction is marked as locally preferred. This policy is enforced within tiles as a memory-aware filtering. If the operands reside in the tile's memory space, the initial (non-relocation) route is used. Otherwise, relocation candidates are evaluated and compared against the conventional data transfer baseline. 

\subsubsection{Firmware Execution} 
The framework assists with reducing global traffic by enabling efficient execution near memory. Compute relocation aligns naturally with hierarchical communication in heterogeneous integration platforms. Examples of instruction execution flows for both conventional data-driven computation and SHIFT are shown in Algorithm~\ref{alg:examples}. 

\begin{algorithm}[t]
    \small
    \caption{Instruction Execution Flows}
    \label{alg:examples}
    \KwData{Instruction $I_k$, Data $D_k$, Data location $N_i$, Source node $N_j$, Candidate node $N_m$}

    \tcp{1: Execution via Data Shifting}
    \If{!(relocation)}{
        \If{$D_k$ not at $N_j$}{
            Route $D_k$ from $N_i$ to $N_j$\;
            Wait until $D_k$ is available at $N_j$\;
        }
        Execute $I_k$ at $N_j$ using $D_k$\;
    }

    \tcp{2: Execution via SHIFT}
    \If{(relocation)}{
        Transmit IIP for $I_k$ from $N_j$ to UC\;
        UC estimates latency to candidate nodes $N_m$\;
        Select destination $N_m$\;
        \If{$N_m \neq N_j$}{
            UC sends \texttt{SHIFT\_TO} to $N_j$\;
            UC sends \texttt{EXECUTE} to $N_m$ with metadata\;
            UC sends \texttt{KILL} to $N_j$\;
        }
        Route $D_k$ from $N_i$ to $N_m$\;
        Wait until $D_k$ is available at $N_m$\;
        Execute $I_k$ at $N_m$ using $D_k$\;
    }
\end{algorithm} 

% Section~\ref{proposed-framework} provided a high-level overview of the proposed framework. Since relocation in some cases can degrade performance, a comprehensive analysis—considering application requirements, detailed topology, communication specifications, and traffic patterns—is necessary to evaluate the effectiveness of the framework, which is presented in Section~\ref{evaluations}. 

\section{Evaluations}\label{evaluations}

\subsection{Experimental Setup}
\label{experimental-setup}

The NoIF, with and without the SHIFT-based routing, is evaluated using cycle-accurate simulations in gem5/Garnet, complemented by a graph-based model that verifies the scaled-out representation of the network results and enables full DSE. FC FLOPS and arithmetic intensity are traced cycle-accurately using Gemmini~\cite{gemmini-dac}, while ChIP additions are derived from RTL implementations. A custom C++ environment ports application traces and dataflows and integrates packaging characteristics into the DSE graph. 
Energy consumption is based on prior measurements of the fine-pitch Si-IF platform and interconnect parameters~\cite{Jangam_2021_tcpmt}, while communication power profiles are obtained from switching activity in the RTL implementation of the UC router and core. The NN model is implemented and trained in PyTorch on a workstation with an NVIDIA RTX 4090 GPU. 

Two sets of configurations are used for the following evaluations: 
(i) A set which is designed to assess the capabilities and effectiveness of the proposed platform under random instruction vectors and data patterns. 
(ii) A set incorporating application-specific considerations for LLMs, including memory configurations and chiplet arrangements. 

\subsubsection{Memory Architecture}
\label{setup-memory-architecture} 

The proposed setup adopts a hybrid memory model combining distributed shared memory (DSM) with HBMs as MCs for large-scale platforms. Each FC include local scratchpads (SPAD) for low-latency, small data access, while UC SPADs serve as tile shared memory. Unlike traditional multi-level caches limited by bandwidth, off-chip DRAM latency, and small SRAMs, the DSM design with fine-pitch HBMs co-located within FC tiles reduces access cycles. In-FC single-cycle SRAM access further enables efficient unstructured sparsity processing~\cite{Lie_2024_micro}. 

Each FC core includes a non-coherent L1, similar to conventional SoCs. UC-managed scratchpads are software-controlled to eliminate coherence traffic and allow relocation to manage data placement. Memory access relies on \texttt{flush}/\texttt{sync} operations at task or relocation boundaries, which provide scalable behavior similar to GPU models \cite{Pal_2019_hpca}. During relocation, dirty data is flushed to the UC or HBM before execution resumes. Table~\ref{table-memory} summarizes the utilized memory architecture. 

\begin{table}[t]
    \noindent\begin{minipage}{\linewidth}
    \centering
    \caption{Memory Architecture and Specifications}
    \label{table-memory}
    \vspace{-2.25mm}
    \begin{tblr}{
      width = \linewidth,
      colspec = {Q[200]Q[550]},
      hline{1-2,7} = {-}{},
      stretch = 0,
    }
    \textbf{Feature} & \textbf{Design strategy}\\
    Architecture & DSM + shared chiplet memory\\
    Access model & \texttt{flush}/\texttt{sync} in tasks and/or relocation\\
    FC scratchpad & 256~KB (SRAM -- baseline configuration in \cite{gemmini-dac})\\
    UC scratchpad & 8~\texttimes~256~KB~=~2~MB (SRAM)\\
    MC (HBM4) & 4-16 stacks -- 16 to 64~GB (BW~=~3.3~TB/s)
    \end{tblr}
    \end{minipage}
    \vspace{-1.5mm}
\end{table}

\subsubsection{Network Architecture} 
\label{setup-network-architecture}

\begin{figure}
    \centering
    \includegraphics[width=\linewidth]{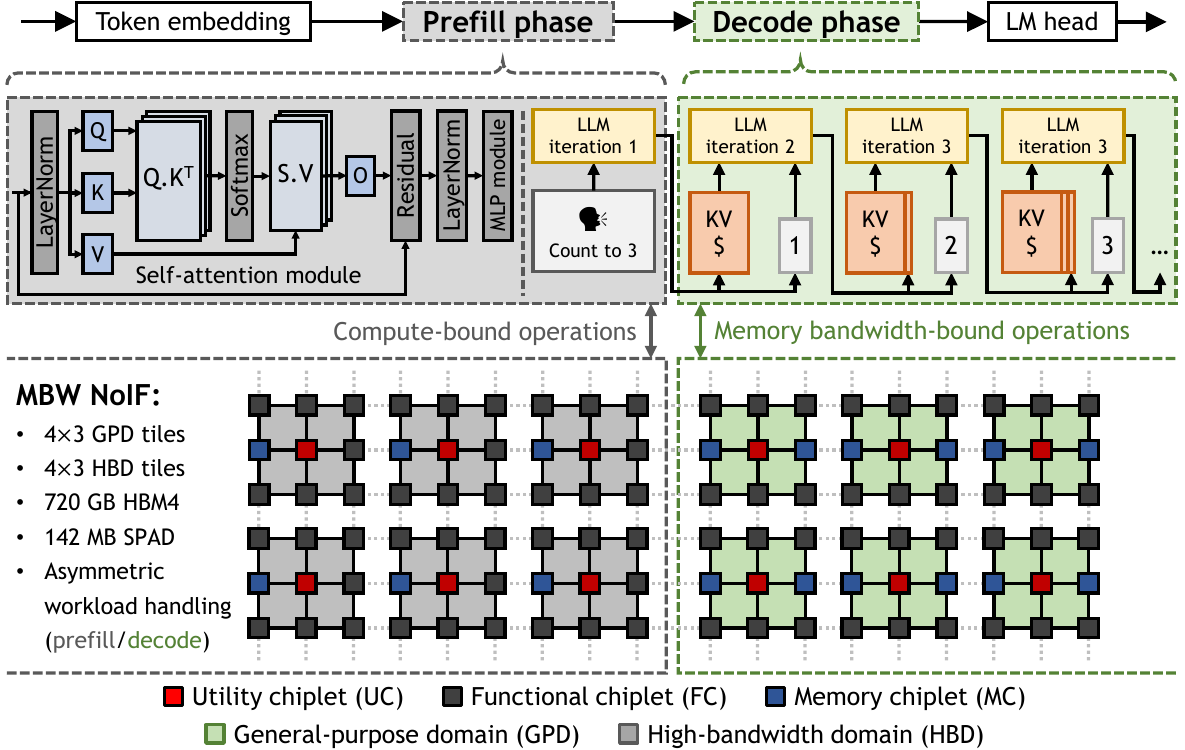}
    \vspace{-6.5mm}
    \caption{Decoder-based LLM inference with asymmetric prefill (compute-bound) and decode (memory bandwidth-bound) mapping into the multi-bandwidth (MBW) network configuration in Table~\ref{table-configs}.}
    \label{fig:mbw-topology} 
\end{figure}

A hierarchical multi-range tile/cluster-based floorplan is adopted to balance communication reach and scalability. UCs are placed in the center of the tiles as the communication hotspot. Each central UC manages a regional domain without overloading the interconnect. 
Additionally, in order to handle the asymmetric requirements of LLM inference, we introduce a multi-bandwidth (MBW) NoIF configuration, where tiles are specialized for either \textbf{prefill} or \textbf{decode} phases, as shown in Fig.~\ref{fig:mbw-topology}. Green regions denote high-bandwidth domains (HBDs) and gray regions represent general-purpose domains (GPDs).

Prefill tiles are compute‑bound, with more FCs and fewer MCs (\textit{i.e.}, GPD), enabling high‑throughput matrix computations. Decode tiles are memory‑bound, incorporating more MCs to support KV‑cache access and bandwidth‑intensive operations (\textit{i.e.}, HBD). Consequently, the MBW configuration is specialized for LLM‑aware asymmetric workload phases.

\begin{figure}
    \centering
    \includegraphics[width=\linewidth]{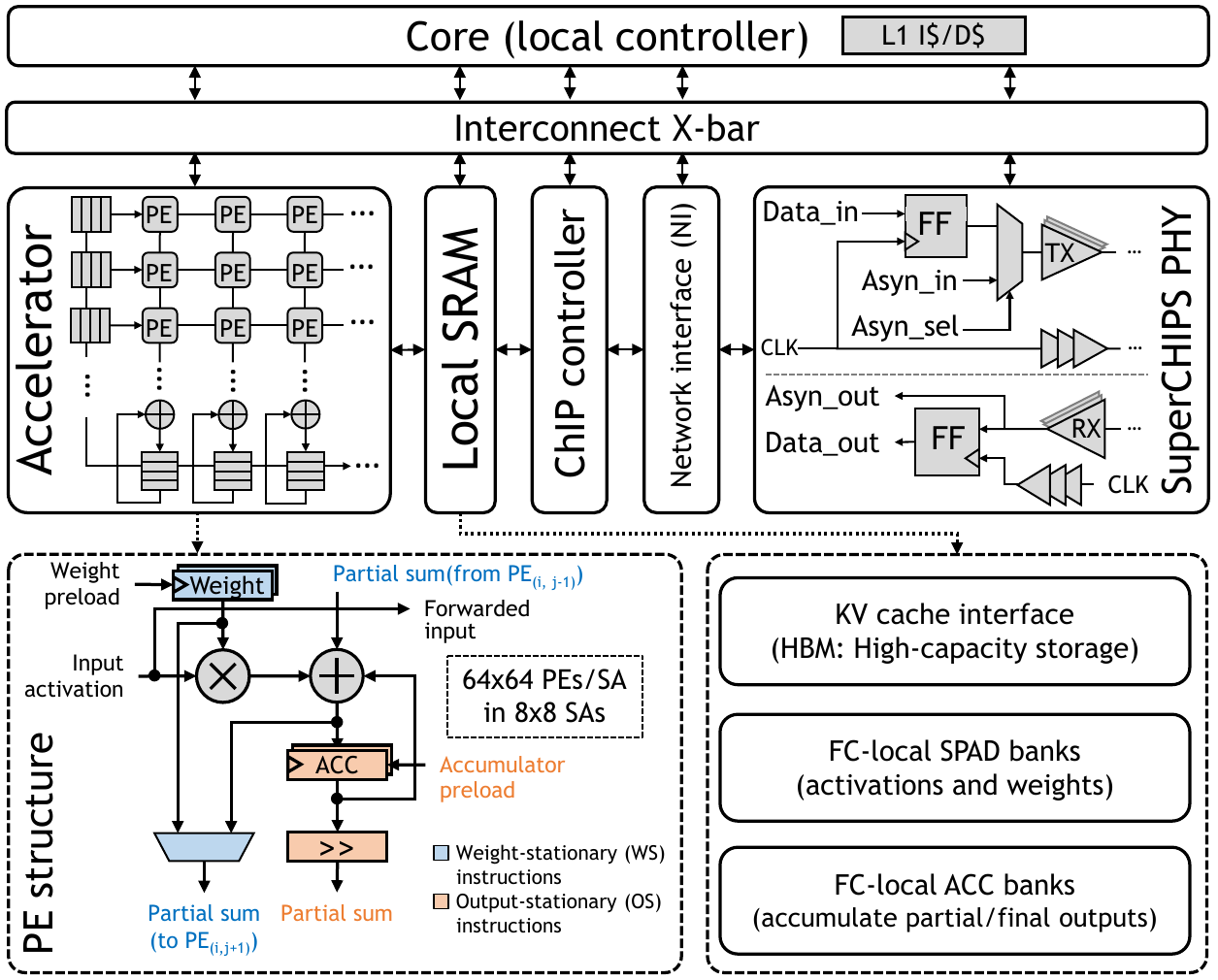}
    \vspace{-7mm}
    \caption{A high-level microarchitecture model of the FC. The PEs in the SA which performs matrix multiplications support both output-stationary (OS) and weight-stationary (WS) dataflows. KV cache data is primarily stored in external HBM4 MCs, which a dedicated interface module and intermediate memory allows for efficient communication.}
    \label{fig:fc-microarchitecture}
    \vspace{-3mm}
\end{figure}

In both HBDs and GPDs, FCs integrate a GEMM accelerator with systolic‑array (SA) processing elements (PEs). The cycle‑accurate behavior of this accelerator is derived from Gemmini \cite{gemmini-dac}, a full‑stack DNN acceleration platform. A high‑level microarchitecture of the modeled FCs is shown in Fig.~\ref{fig:fc-microarchitecture}. 
The SA supports arbitrary read and write access to accumulator entries, with inputs stored in the SPAD and partial sums or outputs maintained in the accumulator (ACC). 

In HBDs, in addition to 2 HBM4 MC-per-tile, UC SPADs are also upgraded to 8~MB, and FCs FP16 arithmetic intensity is increased by 2\texttimes. 
UCs handle both relocation and inter-domain routing, as the router cores maintain access to HBM metadata and the routing history. As a result, global traffic sensing responsibilities are offloaded to these specialized nodes, balancing workloads across domains. 

\begin{table}[t]
    \centering
    \caption{Simulation Parameters}
    \vspace{-2.5mm}
    \noindent\begin{minipage}{\linewidth}
    \label{table-parameters}
    \begin{tblr}{
      width = \linewidth,
      colspec = {Q[100]Q[50]},
      hline{1-2,13} = {-}{},
      stretch = 0, 
    }
    \textbf{Parameter} & \textbf{Value}\\
    Packet/Memory payload width (GPD~$|$~HBD) & 256/256~$|$~512/512~bits\\
    ChIP--SuperCHIPS/SerDes routing latency & 2-4/2-8~cycles\\
    % SerDes routing latency & 2-8~cycles\\
    Clock domain crossing (slower) & 2~cycle\\
    FC/UC/MC controller frequency & 1/1.2/1.8~GHz\\
    Cu pillar pitch/Inter-chiplet spacing & 10~\textmu m/100~\textmu m\\
    Number of interconnect layers & 8 layers\\
    Optimal chiplet/Max. substrate area & 100~mm\textsuperscript{2}/70,685~mm\textsuperscript{2}\\
    Interconnect width/thickness/pitch & 2/2/4 \textmu m\\
    % Technology node & 7~nm FinFET\\
    Interconnect link (asyn.) latency & 503~ps\\
    Link energy-efficiency & 0.15~pJ/bit\\
    Available bandwidth/shoreline & 8~Tb/s/mm
    \end{tblr}
    \end{minipage}
\end{table} 

\subsubsection{ML Dataset and Model}

To validate the approach, a communication-network proxy with comparable per-hop delay behavior under load is used—the Graph Neural Networking Challenge dataset~\cite{graph-neural-networking-challenge}, commonly applied for learning delay and jitter models across topologies and traffic patterns~\cite{10.1145/3314148.3314357}. 

A compact 2-layer ReLU multi-layer perceptron (MLP) ($16\!\rightarrow\!32\!\rightarrow\!1$; ReLU on the hidden layer, linear output) is trained to predict per-neighbor residual latency.
A total of 577 INT8 parameters (\textit{i.e.}, $(16+1){\times}32 + (32+1){\times}1$) were used, which correspond to ($\approx$\,0.58~KB) of weight storage, maintaining a minimal footprint and inference latency.

\subsubsection{Simulation Parameters} 
\label{setup-parameters}

Table~\ref{table-parameters} summarizes the simulation parameters used for evaluation of the proposed framework. A deadlock-free routing mechanism is employed using escape virtual channels (VCs), constrained to operate under a shortest-path adaptive routing policy. A fine-pitch integration substrate is adopted, with ChIP serving as the primary protocol for multi-range communication~\cite{Iyer_2019_ibm_journal, Delavari_2025_ieee_jetcas}. 

\subsubsection{Workload Characterization}
\label{setup-workload}

In the first analysis, random instruction vectors and data patterns are applied to evaluate \textbf{adaptability and scalability under unstructured sparse workloads}. The second analysis executes \textbf{standard LLM benchmarks} on the platform for application-level evaluation. 

\textbf{LLMs:} To assess SHIFT under realistic conditions, we conduct application-driven analysis under LLM workloads, which impose stringent demands on memory-compute interactions and large-scale hardware. LLMs are transformer-based DNNs designed for language understanding and generation through attention and feedforward mechanisms \cite{attention_paper}. 

\textbf{Benchmarks and Datasets:} In this study, decoder-only architectures including LLaMA-2-7B, LLaMA-3-70B, LLaMA-3.1-8B, LLaMA-3.1-70B, Qwen-2-7B, Qwen-3-7B, Falcon-7B, BLOOM-176-B, GPT3-13B and GPT3-175B are selected. Application characterization is performed through context profiling \cite{Zheng_usenix_2022, WaferLLM_2025, Benchmark_paper}, and traces are derived from publicly available inference scenarios used as datasets~\cite{Heo_asplos_2024, huggingface_models}. 

\begin{figure}
    \centering
    \includegraphics[width=\linewidth]{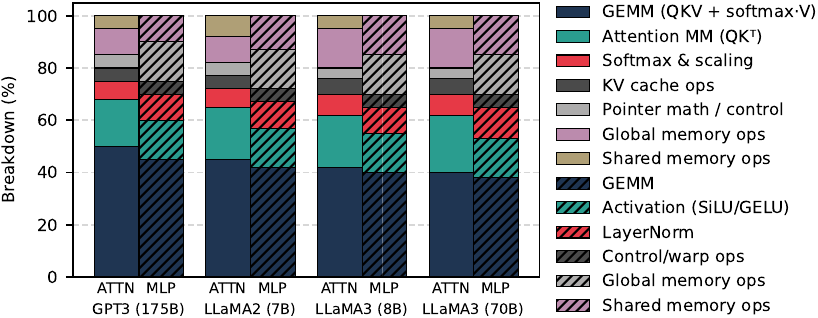}
    \vspace{-7mm}
    \caption{Example of context breakdown in some of the selected LLMs.} 
    \label{fig:profiling} 
\end{figure} 

\textbf{Profiling:} An example breakdown of the context/instruction-per-attention (ATTN) and MLP feedforward blocks, in some of the selected datasets and models, is shown in Fig.~\ref{fig:profiling}. Profiling results shows that data-centric operations—memory access, KV-cache, and GEMM—deliver the largest gains by reducing data movement, averaged in FP16 precision. Normalization and multi-head attention (MHA) provide moderate benefits, while control and scalar instructions show minimal impact, except for LLaMA-2, which benefits from group query attention (GQA) \cite{LLaMA2_paper}. 
In such cases, gains arise mainly from reduced stalls due to low communication demands. Context-aware profiling enables fine-grain modeling by measuring relocation success rates and performance gains within HBDs and GPDs,  using local memory-aware filtering. 

\textbf{Mapping and Dataflow}: 
A weight-stationary (WS) dataflow is employed, where weights are loaded from the bottom edge and retained on-chip during execution. Input activations stream from the left/top edges, while outputs exit from the right/bottom edges. Partial sums are accumulated locally and may be re-streamed or stored. Integrated memory (SPAD/ACC) holds model parameters and runtime state, with overflow data placed in external MCs. Outputs are routed based on the tile/cluster format of IIP and may remain local, be exchanged within the tile, or be relocated via SHIFT.

\subsection{Experimental Results}
\label{results}  

Section~\ref{results-ml} compares the proposed shortest-path algorithm with a lightweight predictive runtime optimization and conventional Dijkstra method. Sections~\ref{results-performance} and~\ref{results-congestion} present results from 10,000 random instruction and data injections under full GPD-based configurations, while Section~\ref{results-application} extends the analysis to the MBW configuration using the LLM workloads from Section~\ref{setup-workload}. 

\subsubsection{ML-Enhanced Routing}
\label{results-ml}

The mean absolute percentage error (MAPE) is used, where $n$ is the sample count, $L_\text{actual}^{(i)}$ is the measured latency, and $L_\text{pred}^{(i)}$ is the predicted latency: 
\begin{equation}
    \label{eq:mape}
    \mathrm{MAPE} = \frac{100}{n}\sum_{i=1}^{n}\left|\frac{L^{(i)}_{\mathrm{actual}} - L^{(i)}_{\mathrm{pred}}}{L^{(i)}_{\mathrm{actual}}}\right|.
\end{equation}
On held-out data, an MAPE of 17.5\% is achieved, supporting the use of $\hat{L}(\cdot)$ for next-hop selection~\cite{7229288, LPNet, Fast-and-Accurate-NoC-Latency-Estimation-for-Application-Specific-Traffics-via-Machine-Learning}. 

Compared to one-shot Dijkstra, the ML policy introduces an online per-hop evaluation, whose cost depends on the UC router architecture. Scalar UCs favor Dijkstra in routing cycles, while INT8 systolic MMUs perform best with the MLP due to efficient matrix–vector operations. A comparison of UC architectural effects on E2E (routing-decision) latency is provided in Table~\ref{tab:uc_arch_latency_cc}. The results include Dijkstra evaluated on both the full network (FullNet) and the reduced subgraph (SubNet, Algorithm~\ref{alg:bidirectional-dijkstra}) as well as the MLP-based method. 

\begin{table}
\noindent\begin{minipage}{\linewidth}
\centering
\footnotesize
\setlength{\tabcolsep}{4pt}
\caption{Normalized E2E Latency Speedup Across UC Architectures}
\vspace{-2.5mm}
\label{tab:uc_arch_latency_cc}
\begin{tblr}{
  width = \linewidth,
  row{2} = {c},
  cell{1}{1} = {r=2}{},
  cell{1}{2} = {c=4}{0.679\linewidth,c},
  cell{3}{2} = {c},
  cell{3}{3} = {c},
  cell{3}{4} = {c},
  cell{3}{5} = {c},
  cell{4}{2} = {c},
  cell{4}{3} = {c},
  cell{4}{4} = {c},
  cell{4}{5} = {c},
  cell{5}{2} = {c},
  cell{5}{3} = {c},
  cell{5}{4} = {c},
  cell{5}{5} = {c},
  hline{1,3,6} = {-}{},
  hline{2} = {2-5}{},
  stretch = 0,
  rowsep = 1.2pt,
}
\textbf{Algorithm} & \textbf{UC router core architecture} &  &  & \\
 & \textbf{Single-core} & \textbf{Multi-cores} & \textbf{Systolic MMU} & \textbf{O3-core}\\
\textbf{FullNet} & 
% 6,534 
2.5$\times$
& 
% 4,665
3.5$\times$
& 
% 6,534 
2.5$\times$
& 
% 4,562
3.4$\times$
\\
\textbf{SubNet} & 
% 2,434 
6.7$\times$
& 
% 1,995 
8.1$\times$
& 
% 2,434
6.7$\times$
& 
% 1,696 
9.5$\times$
\\
\textbf{MLP} & 
% 16,200
1.0$\times$
& 
% 5,232
3.1$\times$
& 
% 720
22.5$\times$
& 
% 2,424 
6.7$\times$
\end{tblr}
\end{minipage}
\vspace{-4mm}
\end{table}

% A greedy, approximate policy is employed; optimality is not guaranteed. Performance depends on training coverage, and feature staleness or domain shift can degrade accuracy. A conservative fallback to Dijkstra may be retained when features are out-of-distribution or when stability checks fail.

\begin{figure}
    \centering
    \includegraphics[width=\linewidth]{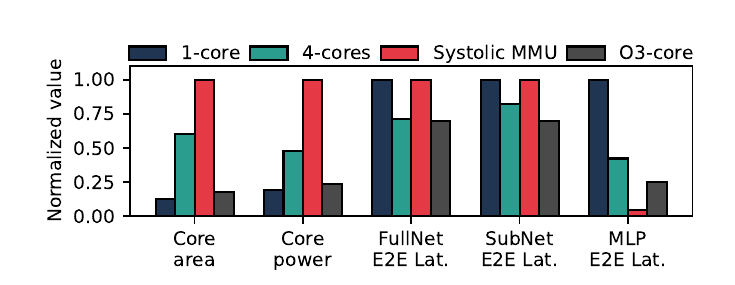}
    \vspace{-9mm}
    \caption{Router performance and hardware cost trade-off with respect to path estimation policy. The systolic MMU with an 8$\times$8 engine incurs the highest hardware cost, while its performance enhancement gap justifies its worthiness for the MLP by efficiently amortizing matrix–vector operations.} 
    \label{fig:trade-off}
\end{figure}

A trade-off between per-hop online evaluation and one-shot planning with respect to UC router core architecture is depicted in Fig.~\ref{fig:trade-off}. Scalar core (simple 6-stage pipelined processors) shows higher decision latency, while multi-core (4-core) and O3 (4-way superscalar) UCs mitigate it moderately. The systolic MMU with an 8$\times$8 engine performs best for the MLP, efficiently amortizing matrix–vector operations. 

\subsubsection{General Performance Analysis}
\label{results-performance}

Depending on data locality and memory access patterns, certain tasks may cause performance degradation instead of improvement. The configurations used for DSE are summarized in Table~\ref{table-configs}, and the corresponding average improvements and degradations across metrics are shown in Table~\ref{table-results}. 

\begin{table}[ht]
    \noindent\begin{minipage}{\linewidth}
    \centering
    \caption{Experimental Setup Configurations} 
    \label{table-configs}
    \vspace{-2.5mm}
    \begin{tblr}{
      width = \linewidth,
      colspec = {Q[97]Q[15]Q[15]Q[15]Q[15]Q[15]Q[15]Q[15]Q[25]},
      column{2-9} = {c},
      hline{1-2,8} = {-}{},
      stretch = 0,
      rowsep = 1.2pt,
    }
    \textbf{Configuration} & \textbf{A} &\textbf{B} & \textbf{I} & \textbf{II} & \textbf{III} & \textbf{IV} & \textbf{V} & \textbf{MBW}\\
    Number of FCs & 7 & 14 & 28 & 63 & 84 & 112 & 252 & 234\\
    Number of UCs & 1 & 2 & 4 & 9 & 12 & 16 & 36 & 36\\
    Number of MCs & 1 & 2 & 4 & 9 & 12 & 16 & 36 & 54\\
    $\sum \text{SRAM}_\text{FC}$ (MB) & 1.8 & 3.6 & 7.2 & 16 & 21.5 & 28.6 & 64 & 142\\
    $\sum \text{SRAM}_\text{UC}$ (MB) & 2 & 4 & 8 & 18 & 24 & 32 & 72 & 180\\
    $\sum \text{HBM4}_\text{MC}$ (GB) & 8 & 16 & 64 & 144 & 192 & 256 & 576 & 720
    \end{tblr}
    \end{minipage}  
\end{table}

\begin{figure*}
    \centering
    \includegraphics[width=\linewidth]{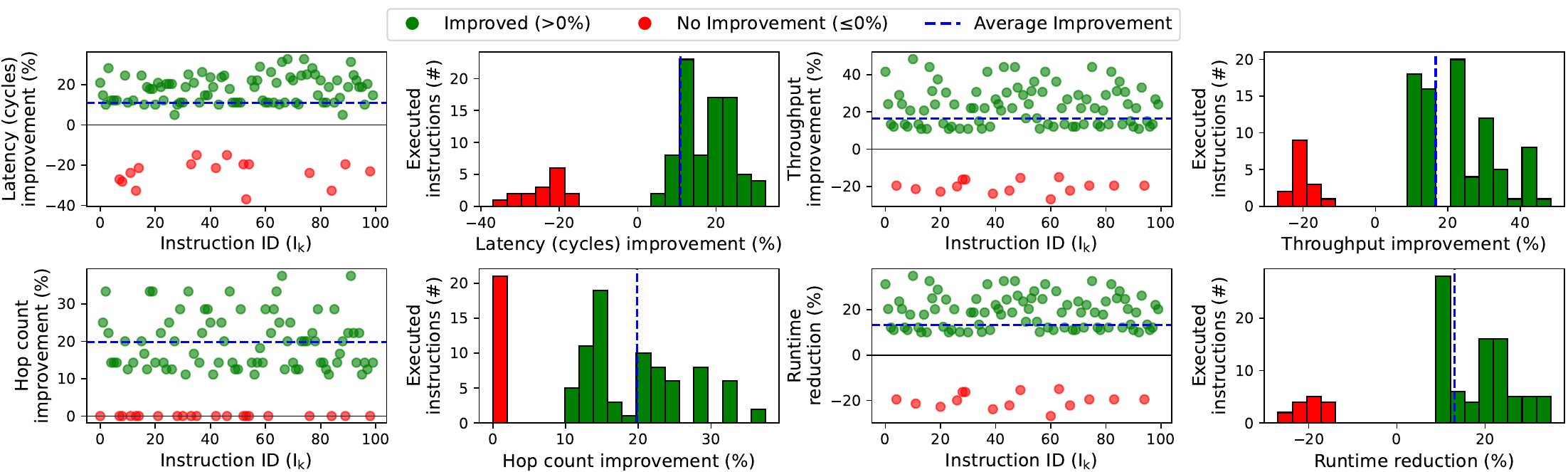}
    \vspace{-7mm}
    \caption{Inter-chiplet latency, hop count, throughput, and runtime (per-kernel, due to heterogeneous clock domains in the system) improvement/degradation analysis for 100 random instructions using SHIFT in configuration~III.} 
    \label{fig:plot-100-analysis} 
    \vspace{-4mm}
\end{figure*}

\begin{table}[ht]
    \caption{Detailed Improvements/Degradations Analysis in E2E Latency and Throughput in Homogeneous Configurations}
    \label{table-results}
    \vspace{-2.5mm}
    \noindent\begin{minipage}{\linewidth}
    \centering
    \footnotesize
    \begin{tblr}{
      width = \linewidth,
      colspec = {Q[350]Q[100]Q[100]Q[100]Q[100]Q[100]Q[100]},
      row{2} = {c},
      row{9} = {c},
      column{3} = {c},
      column{5} = {c},
      column{even} = {c},
      cell{2}{1} = {c=6}{0.933\linewidth},
      cell{9}{1} = {c=6}{0.933\linewidth},
      cell{16}{1} = {c=6}{0.933\linewidth},
      cell{17}{1} = {c=6}{0.933\linewidth},
      cell{18}{1} = {c=6}{0.933\linewidth},
      cell{19}{1} = {c=6}{0.933\linewidth},
      vline{2} = {1,3-9,10-16}{},
      hline{1-3,9-10,16} = {-}{},
      stretch = 0,
      rowsep = 0.9pt,
    }
    \textbf{Metrics} & \textbf{I} & \textbf{II} & \textbf{III} & \textbf{IV} & \textbf{V}\\
    End-to-end latency (cycles) &  &  &  &  &  \\
    Success rate \textsuperscript{1} & 75.2\% & 84.3\% & 88.1\% & 93.8\% & 97.9\%\\
    Avg. improvement \textsuperscript{2} & 16.4\% & 18.0\% & 36.1\% & 53.1\% & 62.5\%\\
    Overall improvement \textsuperscript{3} & 7.4\% & 11.8\% & 32.9\% & 52.6\% & 60.9\%\\
    Max. improvement \textsuperscript{4} & 31.4\% & 33.9\% & 48.7\% & 58.8\% & 76.8\%\\
    Avg. degradation \textsuperscript{2} & \textminus20.5\% & \textminus21.9\% & \textminus39.0\% & \textminus26.8\% & \textminus42.1\%\\
    Max. degradation \textsuperscript{4} & \textminus35.3\% & \textminus39.5\% & \textminus41.1\% & \textminus41.7\% & \textminus50.6\%\\
    Throughput (Tb/s) &  &  &  &  &   \\
    Success rate \textsuperscript{1} & 67.1\% & 77.5\% & 82.9\% & 94.3\% & 97.4\%\\
    Avg. improvement \textsuperscript{2} & 19.8\% & 20.7\% & 49.7\% & 56.2\% & 75.2\%\\
    Overall improvement \textsuperscript{3} & 7.7\% & 11.4\% & 46.1\% & 50.5\% & 71.2\%\\
    Max. improvement \textsuperscript{4} & 37.2\% & 44.2\% & 75.1\% & 65.1\% & 92.0\%\\
    Avg. degradation \textsuperscript{2} & \textminus19.2\% & \textminus20.6\% & \textminus35.4\% & \textminus36.4\% & \textminus41.3\%\\
    Max. degradation \textsuperscript{4} & \textminus34.6\% & \textminus30.9\% & \textminus42.0\% & \textminus44.8\% & \textminus61.5\%\\
    \textsuperscript{1}~Success rate: Ratio of executions successfully relocated that resulted
    in improvements to all instructions. (\#Success(exec.)/$\sum\text{trials}$)~ ~ ~~ &  &  &  &  &   \\
    \textsuperscript{2}~Average improvements/degradations: The average is calculated only over successful/failed relocation trials. &  &  &  &  &   \\
    \textsuperscript{3}~Overall improvement: Total system improvement, accounting for both successful and failed relocation decisions ($\forall~\text{exec.} \in \sum\text{trials}$). &  &  &  &  &   \\
    \textsuperscript{4}~Maximum improvement/degradation: The best/worst of the trials between all executions. &  &  &  &  & 
    \end{tblr}
    \end{minipage}
    \vspace{-2mm}
\end{table} 

\begin{figure}
    \centering
    \includegraphics[width=\linewidth]{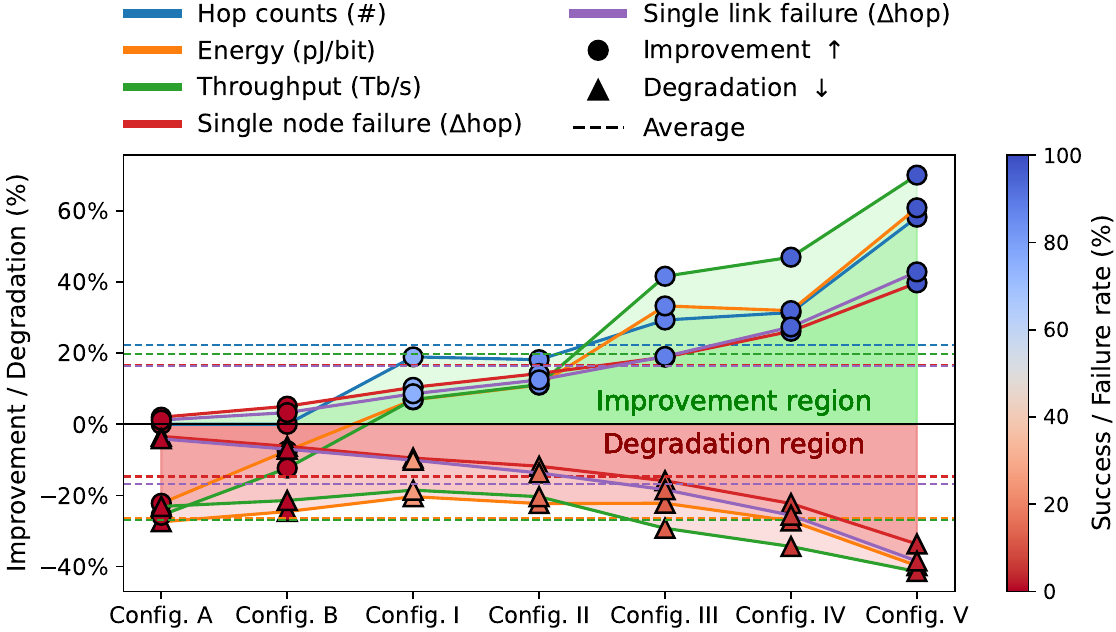}
    \vspace{-6.75mm}
    \caption{Average improvement/degradation trends in hop count, energy-per-bit, throughput, and node-link fault-tolerance ($\Delta$hop) using SHIFT.}
    \label{fig:summary-plot}
\end{figure}

The results in Table~\ref{table-results} and Fig.~\ref{fig:summary-plot} indicate that as network size and workload increase, both the success rate and overall system improvement tend to grow. In contrast, for smaller network sizes (\textit{e.g.}, configurations A and B), the aggregate outcome shows overall degradation, given that the framework introduces \texttt{STALL} states and additional computational overhead. 
\textbf{The evaluations exhibit that for a notable portion of computations, with accordance to their dependencies, a better-positioned node than the pre-assigned will be available at runtime, in the majority of scenarios.}

The trends suggest that the proposed strategy is most suitable for scaled-out architectures rather than small SoCs and workloads. As the network scales, degradation instances also increase; however, the overall improvement continues to rise significantly due to a reduced failure rate (\textit{e.g.}, $<$3\% in configuration~V). 
\textbf{As a result, SHIFT is effective for applications with high compute-memory utilization, where communication cost is a dominant performance factor}. 

Inter-chiplet latency, hop count, throughput, and E2E runtime for 100 random instructions in configuration~III are shown in Fig.~\ref{fig:plot-100-analysis}, highlighting sparse improvements and degradations. The reduced instruction count improves clarity, and histograms depict gain distribution across the system. 

The analysis shows that with the SHIFT framework, configuration~V achieves a maximum throughput improvement of 92\%, with 97.4\% of executions successfully relocated. In addition, configuration~V reduces energy per bit by up to 58.3\%, while configuration~I achieves a 7.6\% reduction. 

\subsubsection{System under Congestion}
\label{results-congestion}

\begin{figure}
    \centering
    \includegraphics[width=\linewidth]{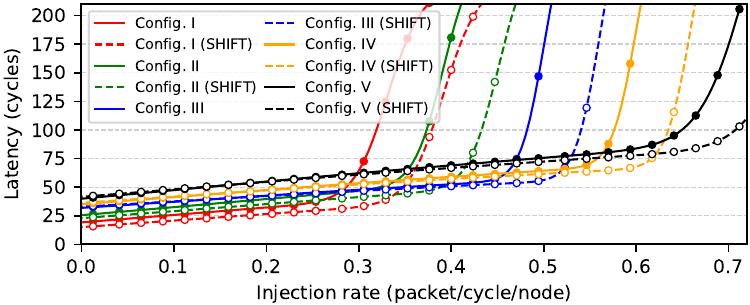}
    \vspace{-7mm}
    \caption{Latency vs. injection rate using SHIFT in Table~\ref{table-configs} configurations.}
    \label{fig:latency-vs-injection}
\end{figure}

Platform congestion stress is evaluated under varying packet injection rates using random GEMM kernels with sparse data sources to ensure fair evaluation. The simulation includes 10,000 warm-up cycles followed by a 100,000-cycle evaluation period. Fig.~\ref{fig:latency-vs-injection} illustrates latency versus injection rate for both with and without SHIFT.

\begin{figure*} 
    \centering
    \includegraphics[width=\linewidth]{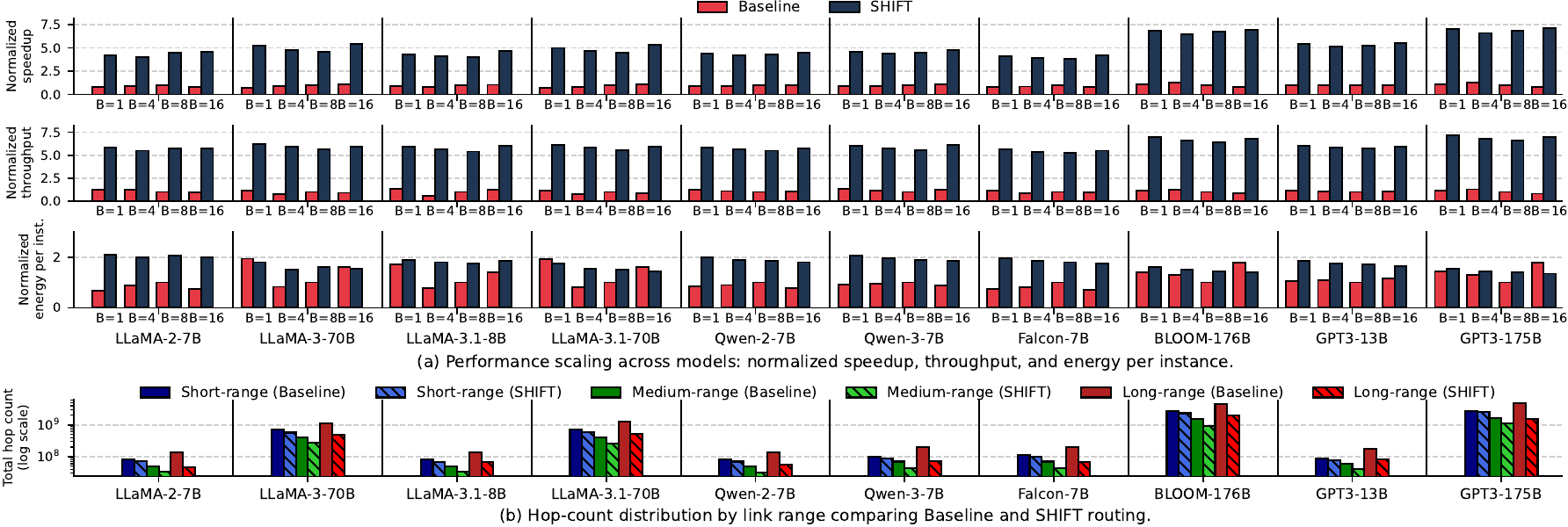}
    \vspace{-7.25mm}
    \caption{SHIFT improvements against baseline NoIF in: (a) normalized runtime/throughput/energy vs. batch size (denoted as B), and (b) Hop count per link-range for the selected LLM benchmarks. The baseline bars represent the results on the same MBW configuration without the SHIFT framework.}
    \label{fig:application-runtime}
    \vspace{-4.75mm}
\end{figure*}

In AI workloads, the injection rate typically ranges from 0.2 to 0.5 during inference due to intensive data movement in GEMM operations. During training, especially in distributed setups, rates can be higher, sometimes reaching 0.7 or more, driven by gradient exchanges and synchronization. Actual rates depend on factors like model size, batch size, and communication specifications. \textbf{As the saturation point shifts higher with the SHIFT framework, the results suggest suitability for congestive workloads such as AI and HPC}. 

\subsubsection{Application-Level Enhancements} 
\label{results-application} 

LLMs serve as a strong example of large-scale applications with significant computational and memory demands. The MBW configuration is used as the testbed for these evaluations. Average improvements in runtime, throughput (tokens/sec), and energy efficiency (pJ/operation) with and without SHIFT, across varying batch sizes (B), are reported for selected benchmarks in Fig.~\ref{fig:application-runtime}.~(a), all normalized to the baseline B=8 for comparison. In addition, changes in hop counts by link type and utilization in MBW NoIF are broken down and compared in Fig.~\ref{fig:application-runtime}.~(b). 

\begin{figure}
    \centering
    \includegraphics[width=0.95\linewidth]{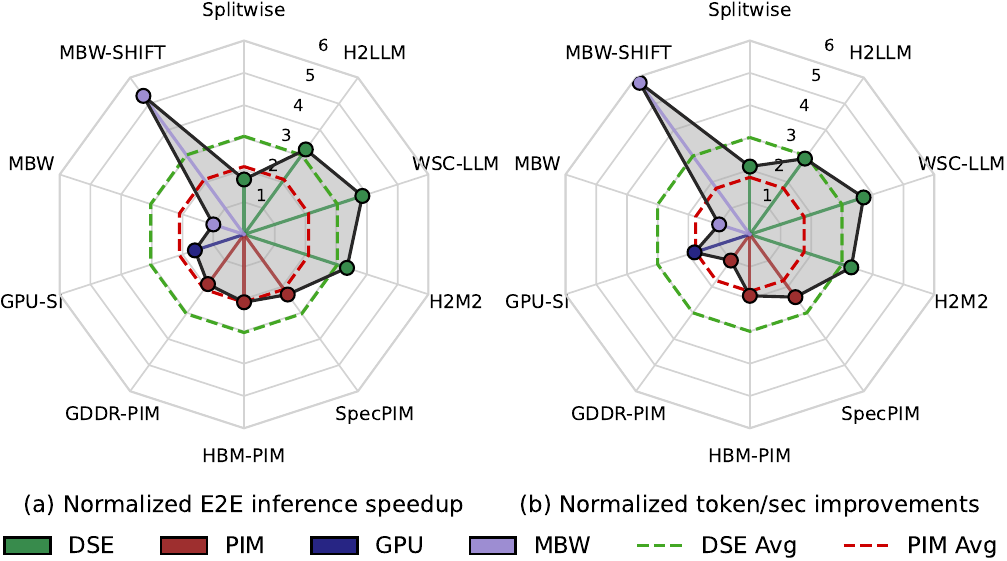}
    \vspace{-2.5mm}
    \caption{Speedup and application throughput comparison with SOTA chiplet-based and wafer-scale LLM services. All evaluations are based on LLaMA-3 (8B and 70B) models and normalized to the baseline MBW configuration.}
    \label{fig:sota-llm-comparison}
    \vspace{-1mm}
\end{figure}

In addition, a comparison of average E2E latency speedup against SOTA LLM services for wafer-scale architectures is shown in Fig.~\ref{fig:sota-llm-comparison}. A general comparison of some of these platforms and some other similar work are also shown in Table~\ref{table-llm-platforms}. This includes speculative inference on GPU, PIM-based approaches such as Samsung HBM-PIM~\cite{HBM_PIM}, SK-Hynix PIM solutions (GDDR-PIM)~\cite{GDDR6_PIM}, and SpecPIM~\cite{SpecPIM} on an A100 GPU host~\cite{Choquette_2021_ieee_micro_nvidia}. 
We also consider WSC-LLM~\cite{Xu_isca_2025_llm} and Theseus \cite{Zhu_tcad_2025_llm}, architecture–scheduling co-exploration frameworks; H\textsuperscript{2}LLM~\cite{H2-LLM}, a hybrid-bonding-based heterogeneous accelerator; Splitwise~\cite{Azure_Splitwise}, which partitions LLM inference phases across machines; and H2M2~\cite{H2M2}, a hardware-driven heterogeneous memory management co-design, as other DSE approaches for LLM inference. 
Apart from LLM services, other SOTA large-scale network-level studies are also compared in Table~\ref{table-llm-platforms}, including FRED, which performs network-level interconnect optimization~\cite{Rashidi_2025_isca}, and Gemini, which demonstrates DSE for DNN workload mappings~\cite{Cai_2024_hpca}.

\begin{table}
\noindent\begin{minipage}{\linewidth}
    \centering
    \footnotesize 
    \caption{SOTA Chiplet-Based and Wafer-Scale Services}
    \label{table-llm-platforms}
    \vspace{-2.75mm}
    \begin{tblr}{
      width=\linewidth,
      colspec={Q[85]Q[37.5]Q[90]Q[85]Q[70]Q[45]},
      column{2}={c},
      column{3}={c},
      column{4}={c},
      column{5}={c},
      column{6}={c},
      hline{1,2,11,12}={-}{},
      stretch=0,
      % rowsep=1.0pt
    }
    \textbf{Name} & \textbf{Ref.} & \textbf{Platform} & \textbf{Packaging} & \textbf{Topology} & \textbf{Hetero.}\\
    Splitwise & \cite{Azure_Splitwise} & GPU & Monolithic & DC-net & No \\
    SpecPIM & \cite{SpecPIM} & HBM & 3D (TSV) & 2D-mesh & No \\
    WSC-LLM & \cite{Xu_isca_2025_llm} & Chiplet & Interposer & 2D-mesh & No \\
    Theseus & \cite{Zhu_tcad_2025_llm} & WSC & RDL & 2D-mesh & Yes \\ 
    H\textsuperscript{2}LLM & \cite{H2-LLM} & Chiplet & HB & 2D-mesh & Yes \\
    H2M2 & \cite{H2M2} & Interposer & N/A & 2D-mesh & Yes \\
    FRED & \cite{Rashidi_2025_isca} & NPU & Si-IF & Clos & No \\
    Gemini & \cite{Cai_2024_hpca} & Chiplet & Interposer & 2D-mesh & No \\
    DAC'21 & \cite{Pal_dac_2021} & WSC/chiplet & Si-IF & 2D-mesh & No \\
    SHIFT & -- & WSC/chiplet & Si-IF & NoIF & Yes \\
    \SetCell[c=6]{l} WSC: Wafer-scale -- HB: Hybrid bonding -- RDL: Re-distribution layer\\
    \end{tblr}
\end{minipage}
\vspace{-1mm}
\end{table}

As a result, average normalized improvements of 4.9$\times$, 5.9$\times$, and 1.8$\times$ are achieved in runtime speedup, throughput, and energy-efficiency, respectively. Furthermore, in comparisons with SOTA LLM services and compute–memory co-design approaches \cite{HBM_PIM, GDDR6_PIM, SpecPIM, Xu_isca_2025_llm,H2-LLM, Azure_Splitwise, H2M2}, as shown in Fig.~\ref{fig:sota-llm-comparison}, the proposed framework outperforms PIM and DSE averages by 74.5\% and 152.4\%, respectively. 

\textbf{At the firmware-level, SHIFT impact extends beyond memory or GEMM operations, as the relocation framework applies to all instruction types, unlike domain- and application-specific enhancements in SOTA platforms. Thus, this strategy can be applied to a wider range of applications, still delivering substantial improvements.} 

\subsubsection{SOTA Large-Scale Architectures}

To evaluate the contribution of the proposed strategy to SOTA large-scale computing platforms, a comparison is conducted against Cerebras wafer-scale engine (WSE-3)~\cite{Lie_2024_micro}, Tesla Dojo ExaPOD~\cite{Talpes_2023_micro}, and NVIDIA H100~\cite{Choquette_2023_ieee_micro_nvidia} in Table~\ref{table-accelerator-comparison}. 

\begin{table}[ht]
\noindent\begin{minipage}{\linewidth}
\centering
\caption{Comparison of Large-scale AI Accelerators}
\label{table-accelerator-comparison}
\vspace{-2.5mm}
\footnotesize  
\begin{tblr}{
width=\linewidth,
colspec={Q[170]Q[90]Q[90]Q[110]Q[90]Q[90]},
column{2}={c},
column{3}={c},
column{4}={c},
column{5}={c},
column{6}={c},
hline{1,2,7}={-}{},
stretch=0.25,
}
\textbf{Performance metrics} & \textbf{MBW (base.)} & \textbf{MBW (SHIFT)} & \textbf{Cerebras WSE-3} & \textbf{Tesla Dojo} & \textbf{DGX H100} \\
No. of cores & 234* & 234* & 900,000 & $>\text{10}^\text{6}$ & 116,736\\
Process (nm) & 22~FDX & 22~FDX & 5~TSMC & 7~TSMC & 4~TSMC \\
W/mm$^\text{2}$ & 0.27 & 0.25 & 0.65 & 0.35 & 1.56 \\
FP16 PFLOPS & 122.7 & 149.7 & 125 & 1080 & 15.8 \\
GFLOPS/W/core & 55.2 & 72.5 & 0.01 & 0.07 & 0.01 \\
\SetCell[c=6]{l} *~Number of FCs (each include 8\texttimes8 SAs with 64\texttimes64 PEs).\\
\end{tblr}
\end{minipage}
\end{table}

Cerebras WSE-3 integrates 900,000 AI cores with 44~GB on-chip SRAM, occupying 46,255~mm\textsuperscript{2} and reporting peak power of 23-30~kW, while Tesla Dojo ExaPOD adopts a different model with the 645~mm\textsuperscript{2} D1 tiles and 900~GB/s memory bandwidth. The NVIDIA H100 integrates 80~GB of HBM3 with NVLink connectivity and 14,592 CUDA cores, while DGX H100 is a complete AI server system with 8\texttimes~H100 GPUs plus CPUs, networking, NVSwitch fabric, and storage. The proposed MBW NoIF configuration occupies 35,310~mm\textsuperscript{2} and has a peak power of 9.5~kW in GF~22FDX process, which can be further reduced by a $\sim$650-700~W through SHIFT. 

\begin{figure}
    \centering
    \includegraphics[width=\linewidth] {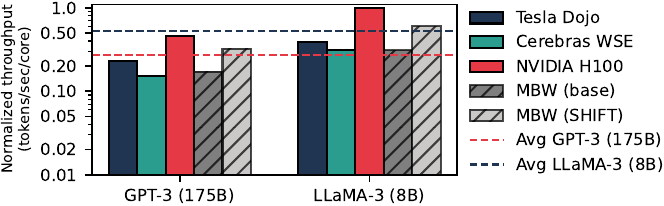} 
    \vspace{-7.25mm} 
    \caption{Comparison of normalized throughput-per-core across GPT3-175B and LLaMA-3-8B~\cite{GPT3_paper, LLaMA3_paper} on SOTA large-scale platforms \cite{Talpes_2023_micro, Lie_2024_micro, Choquette_2023_ieee_micro_nvidia, Prasanna_HPEC_2024, Zhu_tcad_2025_llm}, and the MBW NoIF with and without SHIFT framework.} 
    \label{fig:application-comparison}
\end{figure}

While WSE-3 and Dojo provide greater computational resources, the normalized per-core performance (Fig.~\ref{fig:application-comparison} and Table~\ref{table-accelerator-comparison}), along with the area and power gaps, highlights the scalability and efficiency of runtime relocation. 
Due to simulation constraints, scaling to the core counts of these platforms is infeasible; therefore, throughput is measured per core (\textit{i.e.}, tokens/sec/core) and normalized for comparison. 
Although the MBW platform uses fewer cores than SOTA counterparts and initially delivers lower average throughput, SHIFT enables it to surpass their benchmark averages. 

Cerebras achieves scale through distributed SRAM within its cores, but this design limits adaptability to heterogeneous workloads. Dojo instead uses a traditional NoC-based organization, which suffers from long-distance communication overhead as system size grows. These constraints prevent wafer-scale architectures from sustaining performance on heterogeneous workloads with compute-heavy and memory-bound operations \cite{Zhu_tcad_2025_llm}. 
As a result, the MBW NoIF achieves a power dissipation per unit area of 0.27~W/mm\textsuperscript{2}, which decreases to 0.25~W/mm\textsuperscript{2} with SHIFT, compared to 0.65, 0.35, and 1.56~W/mm\textsuperscript{2} for WSE-3, Dojo, and DGX H100 respectively. 

\begin{table}[ht]
\noindent\begin{minipage}{\linewidth}
    \centering
    \caption{Power Breakdown based on Chiplets and Relocation Stages, with and without SHIFT Framework}
    \label{table-power-breakdown}
    \vspace{-2.5mm}
    \resizebox{\columnwidth}{!}{
    \begin{tblr}{
      width = \linewidth,
      colspec = {Q[70]Q[50]Q[50]Q[160]Q[35]},
      row{1} = {c},
      column{2} = {c},
      column{3} = {c},
      column{5} = {c},
      cell{1}{1} = {c=5}{\linewidth, c},
      cell{2}{1} =  {c=5}{\linewidth, c},
      cell{10}{1} = {c=5}{\linewidth},
      cell{11}{1} = {c=5}{\linewidth},
      vline{4} = {1-9}{},
      hline{1-4,10} = {-}{},
      stretch=0,
    }
    \textbf{Baseline power dissipation per unit area: 0.27~W/mm\textsuperscript{2}}\\\textbf{SHIFT power dissipation per unit area: 0.25~W/mm\textsuperscript{2}} &  &  &  & \\
    \textbf{Chiplets} & \textbf{Baseline\textsuperscript{1}} & \textbf{SHIFT\textsuperscript{2}} & \textbf{SHIFT stage} & \textbf{Share}\\
    HBD FCs & 37.8\% & 35.8\% & IIP Generation (FC) & 1.41\%\\
    GPD FCs & 26.0\% & 27.4\% & Buffering + decoding & 3.83\%\\
    HBD UCs & 1.3\% & 1.6\% & Shortest path estimation & 7.54\%\\
    GPD UCs & 0.8\% & 1.1\% & CMD dispatch & 4.31\%\\
    HBD MCs & 22.7\% & 22.6\% & Routing and execution & 71.83\%\\
    GPD MCs & 11.4\% & 11.5\% & Write-back/commit + idle & 11.08\%\\
    \textsuperscript{1}~Portion (\%) in total power consumption without SHIFT.\\
    \textsuperscript{2}~Portion (\%) in total power consumption with SHIFT.~ &  &  &  & 
    \end{tblr}
    }
\end{minipage}
\vspace{-1mm}
\end{table}

\begin{figure}[ht]
    \centering
    \includegraphics[width=\linewidth]{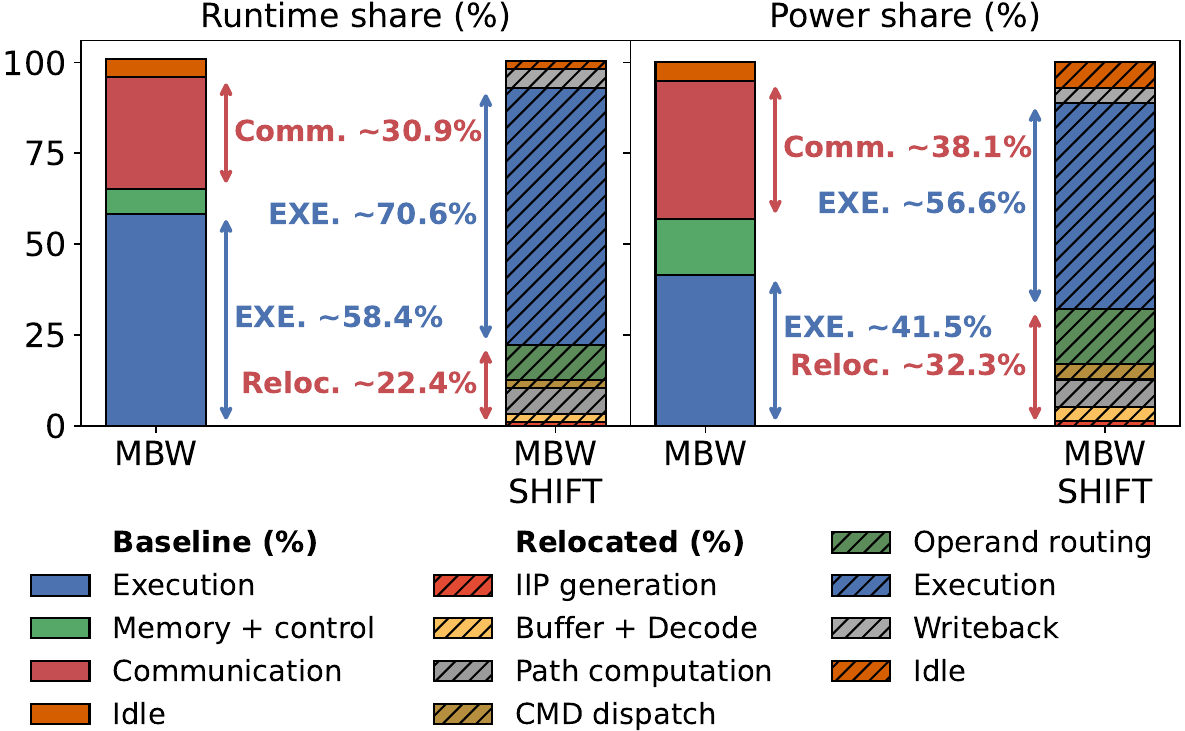}
    \vspace{-6mm}
    \caption{Power and runtime breakdown and comparison.}
    \label{fig:breakdown}
\end{figure}

Table~\ref{table-power-breakdown} provides the power consumption breakdown by chiplet types, LLM inference context, and relocation stages, with the distribution changes reflecting the workload-balancing impact of SHIFT and showing reductions across all context classes. 
Runtime and power changes breakdowns during execution are shown in Fig.~\ref{fig:breakdown}, as and evidence of the SHIFT effect on communication-level savings. 

\section{Conclusions}\label{conclusions}

A novel topology-agnostic framework for runtime communication-aware relocation of compute nodes--rather than deterministic routing toward the corresponding node in the network--was presented in this paper. \textbf{The proposed strategy, due to its scalability, is designed as a system-level methodology for workload-aware runtime optimization in wafer-scale and heterogeneous architectures}. Applications with sparse data dependencies or dynamic scheduling demands benefit the most from runtime resource reallocation, highlighting the suitability of SHIFT for data-center-scale AI inference. The proposed strategy is independent of the underlying platform and can potentially be extended to other applications, representing a future direction of this work. 

The results exhibit average improvements of 16.4\%-62.5\% in latency, 19.8\%–75.2\% in throughput, and up to 58.3\% in energy-efficiency, with success rates ranging from 67.1\% to 97.4\% across various configurations under unstructured traffic.

Furthermore, when implemented on a heterogeneous configuration (MBW NoIF), the proposed framework achieves average normalized improvements of 4.9\texttimes, 5.9\texttimes, and 1.8\texttimes in runtime, throughput, and energy-efficiency, respectively, across variants of LLaMA-2, LLaMA-3, LLaMA-3.1, Qwen-2, Qwen-3, Falcon, BLOOM, and GPT-3, while outperforming SOTA wafer-scale and chiplet-based LLM services. 

\textbf{The key enabler of these improvements is the efficient use of UCs and hierarchical multi-range routing, which offload network management from compute cores to a communication-aware architectural layer.}

The SHIFT framework is broadly comparable to approaches such as PIM/PNM, application-architecture co-design, and domain-specific accelerators. \textbf{SHIFT is the most scalable solution for heterogeneous integration co-optimization, offering lower hardware cost and complexity while supporting adaptive workload placement for data transfers, matrix, and scalar operations.} In contrast, PIM primarily affects limited arithmetic and certain memory access operations, whereas scheduling application are entirely application- and input-dependent, which both approaches can serve as complementary components to the SHIFT. 

\textbf{Overall, the SHIFT framework provides a robust foundation for scalable runtime optimization, delivering significant gains in latency, throughput, and hardware efficiency while offering full flexibility for heterogeneous workloads and architectures.} 
Additionally, since the impact of the framework depends on data structure and placement, SHIFT is expected to deliver similar gains in data-center workloads–a key direction for future evaluation. Future improvements in multi-cycle relocation prediction techniques and application-aware co-scheduling can further boost performance, positioning this methodology as a promising solution for next-generation computing platforms.  

\bibliographystyle{myIEEEtran}
\bibliography{refs}

\balance
\end{document}